\documentclass[10pt,journal,twocolumn,twoside]{IEEEtran} 
\usepackage{xpatch}
\usepackage{graphicx}
\usepackage{subfigure}
\usepackage{epstopdf}
\usepackage{float}
\usepackage{algorithmic}
\usepackage{array}
\usepackage{amsmath}
\usepackage{amssymb}
\usepackage{mdwmath}
\usepackage{mdwtab}
\usepackage{eqparbox}
\usepackage{stfloats}
\usepackage{fixltx2e}
\usepackage{tabularx}
\usepackage{cases} 
\usepackage{booktabs}
\usepackage{threeparttable}
\usepackage{xcolor}
\usepackage{makecell}
\usepackage{multirow}
\usepackage{flushend}
\usepackage[boxed,ruled,commentsnumbered]{algorithm2e}
\usepackage{url}
\usepackage{cite}
\usepackage{etoolbox}
\makeatletter
\patchcmd{\@makecaption}
  {\scshape}
  {}
  {}
  {}
\makeatletter
\patchcmd{\@makecaption}
  {\\}
  {.\ }
  {}
  {}
\makeatother
\def\tablename{Table}
\allowdisplaybreaks[4]

\newtheorem{remark}{Remark}

\newcommand{\diag}{\mathop{\mathrm{diag}}}
\newcommand{\tr}{\mathop{\mathrm{tr}}}

\makeatletter

\newcommand{\Rmnum}[1]{\expandafter\@slowromancap\romannumeral #1@}
\makeatother

\newcounter{MYtempeqncnt} 

\makeatletter
\renewcommand*{\@opargbegintheorem}[3]{\trivlist
      \item[\hskip \labelsep{\bfseries #1\ #2}] \textbf{(#3):}\ }
\makeatother     

\begin{document}

\makeatletter
\def\changeBibColor#1{%
  \in@{#1}{}
  \ifin@\color{red}\else\normalcolor\fi
}
 
\xpatchcmd\@bibitem
  {\item}
  {\changeBibColor{#1}\item}
  {}{\fail}
 
\xpatchcmd\@lbibitem
  {\item}
  {\changeBibColor{#2}\item}
  {}{\fail}
\makeatother

\title
{
Unified Integrated Sensing and Communication Signal Design: A Sphere Packing Perspective
}
\author{Shuaishuai Guo,~\IEEEmembership{Senior Member, IEEE}, and Kaiqian Qu
\thanks{The work is supported in part by the National Natural Science Foundation of China under Grant 62171262; in part by Shandong Provincial Natural Science Foundation under Grant ZR2021YQ47; in part by the Taishan Young Scholar under Grant tsqn201909043; in part by Major Scientific and Technological Innovation Project of Shandong Province under Grant 2020CXGC010109. (\emph{*Corresponding author: Shuaishuai Guo}).}
\thanks{S. Guo and K. Qu are with the School of Control Science and Engineering, and also with Shandong Key Laboratory of Wireless Communication Technologies, Shandong University, Jinan 250062, China (e-mail: shuaishuai$\_$guo@sdu.edu.cn; qukaiqian@mail.sdu.edu.cn). }
   }
\maketitle


\begin{abstract} 
The design of communication signal sets is fundamentally a sphere packing problem. It aims to identify a set of $M$ points in an $N$-dimensional space, with the objective of maximizing the separability of points that represent different bits. In contrast, signals used for sensing targets should ideally be as deterministic as possible. This paper explores the inherent conflict and trade-off between communication and sensing when these functions are combined within the same signal set. We present a unified approach to signal design in the time, frequency, and space domains for integrated sensing and communication (ISAC), framing it as a modified sphere packing problem. Through adept formula manipulation, this problem is transformed into a large-scale quadratic constrained quadratic programming (QCQP) challenge. We propose an augmented Lagrangian and dual ascent (ALDA) algorithm for iterative problem-solving. The computational complexity of this approach is analyzed and found to be daunting for large, high-dimensional signal set designs. To address this, we introduce a bit-dimension-power splitting (BDPS) method. This method decomposes the large-scale QCQP into a series of smaller-scale problems that can be solved more efficiently and in parallel, significantly reducing the overall computational load. Extensive simulations have been conducted to validate the effectiveness of our proposed signal design methods in the context of ISAC.

\end{abstract}


\section{Introduction} 
\IEEEPARstart{I}{ntegrated} sensing and communication (ISAC) is poised to be a key technology in sixth-generation (6G) communication networks, offering ubiquitous sensing capabilities  \cite{8999605,10063187}. ISAC enhances spectrum and energy efficiency by sharing the same signal waveform for both communication and sensing. However, achieving the optimal balance between these functions presents a significant challenge. {For radar sensing, deterministic signals are typically required to achieve optimal sensing performance \cite{9771849}. These signals are optimized according to specific criteria, such as minimizing integrated sidelobes. With deterministic waveforms, the radar receiver can achieve sensing objectives by detecting changes in certain characteristics of the received signals. However, communication signals need to be as random (separable) as possible to carry information. This randomness (separable) makes the signals non-stationary, which is not conducive to radar sensing.} This paper delves into the conflict and trade-off inherent in integrating communication and sensing within the same signal set.

In communication, signal set design is essentially a sphere packing problem, focusing on densely packing spheres without overlap in an $N$-dimensional space. This is mathematically formulated as finding $M$ points (representing $m=\log_2 M$ bits) in an $N$-dimensional space, maximizing the Euclidean distance between the closest pair under a given average power constraint $P$, i.e., \cite{Agrell2014}
\begin{align}
(\textbf{P1}):~\mathrm{Given}: &~M,N,P\notag\\
\mathrm{Find}:&~\mathcal{X}=\{\mathbf{x}_1,\mathbf{x}_2,\cdots,\mathbf{x}_M\}\notag\\
\mathrm{Maximize}:&~\min_{\forall k,l,k\neq l} ||\mathbf{x}_k-\mathbf{x}_l||_2\\
\mathrm{Subject~to}:&~\frac{1}{M}\sum_{k=1}^{M}||\mathbf{x}_k||_2^2\leq P\notag.
\end{align}
{By solving \textbf{P1}, it is possible to minimize the communication symbol error rate (SER) and maximize the communication mutual information (MI) in the absence of channel state information (CSI).

For sensing applications, it is desirable to use optimized and deterministic
waveforms, with desirable properties like minimum integrated sidelobe level (ISL), desired spatial directivity pattern or optimal ambiguity function (AF) \cite{he_li_stoica_2012}. While there are various signal waveform designs for active sensing systems \cite{he_li_stoica_2012}, this paper assumes the availability of an reference sensing signal\footnote{For example, the linear frequency modulated (LFM) waveform is the most commonly used time-domain waveform in traditional radar systems due to its favorable ambiguity characteristics. In some ISAC research, it is used as a reference time-domain sensing waveform \cite{2021JointTW}. Another example is in MIMO radar, where it is often necessary to optimize spatial waveforms to accommodate different sensing requirements. Specifically, when searching a particular area, the spatial waveform typically needs to exhibit a wide beam characteristic. Conversely, when tracking a specific target, the spatial waveform needs to have a narrow beam characteristic. At this point, such waveforms optimized specifically for radar function can serve as spatial reference sensing waveforms in ISAC, as illustrated in the literature \cite{8386661}. }}, represented by $\mathbf{x}_0$. The task is to find a signal set that fulfills both communication and sensing functions, formulated as a sphere-packing problem with similarity constraint:
\begin{align}
(\textbf{P2}):~\mathrm{Given}: &~M,N,P,\mathbf{x}_0,\epsilon\notag\\
\mathrm{Find}:&~\mathcal{X}=\{\mathbf{x}_1,\mathbf{x}_2,\cdots,\mathbf{x}_M\}\notag\\
\mathrm{Maximize}:&~\min_{\forall k,l,k\neq l} ||\mathbf{x}_k-\mathbf{x}_l||_2\\
\mathrm{Subject~to}:&~\frac{1}{M}\sum_{k=1}^{M}||\mathbf{x}_k||_2^2\leq P,\notag\\
&~||\mathbf{x}_k-\mathbf{x}_0||_2\leq \epsilon, \forall k\notag,
\end{align}
{where $\epsilon$ is the sensing waveform similarity tolerance}. This paper aims to develop universal solutions for this fundamental problem, facilitating its application in ISAC systems
\subsection{Related Works on ISAC Signal Design}
The design of ISAC signals, crucial to ISAC technology, has seen extensive study. Current research divides ISAC waveform/signal design into two categories: designs using non-overlapping resources and those employing fully unified signal designs that occupy overlapping resources \cite{liufan2022}. 

Earlier studies have predominantly focused on ISAC waveform designs that utilize non-overlapping resources, primarily due to their simplicity and ease of implementation \cite{liufan2022,Han2013,9728752,8094973,9049009,9078732,9771849,6503914,7814210}. This approach typically involves allocating orthogonal resources across multiple dimensions such as time \cite{Han2013,9728752}, frequency \cite{8094973,9049009,9078732,9771849}, and space \cite{6503914,7814210}. A notable example in the time domain is presented in \cite{Han2013}, where a duplex radar-communication waveform design is proposed. This design divides the transmission duration into distinct radar and communication cycles. Radar cycles utilize a trapezoidal frequency modulated continuous wave (FMCW), comprising an up-chirp, a constant-frequency period, and a down-chirp. Communication cycles, on the other hand, can employ any modulation method. These cycles alternate to prevent time-domain interference, a concept that has been standardized in IEEE 802.11ad \cite{9728752}. In the frequency domain, studies like \cite{8094973,9049009,9078732,9771849} allocate sensing and communication functionalities to different subcarriers, considering channel conditions, functional requirements, and transmitter power budgets. Meanwhile, spatial-domain approaches \cite{6503914,7814210} transmit sensing and communication waveforms over different spatial beams or into each other's null space to avoid interference. While these methods are straightforward and readily implementable due to their decoupling of communication and sensing functions, they exhibit a key drawback: inefficient utilization of time, spectrum, and space, leading to reduced energy efficiency.

The challenge of inefficient resource utilization in ISAC has led to a growing interest in fully unified ISAC signal designs that utilize overlapping resources. These can be broadly categorized into three approaches: 
\begin{itemize}
    \item Communication-centric design (CCD): This approach primarily enhances conventional communication waveforms with sensing capabilities. For instance, \cite{5776640,9729203} explore integrated signal designs in multicarrier communication systems by modifying orthogonal frequency division multiplexing (OFDM) signals. Sturm and Wiesbeck in \cite{5776640} proposed a radar processing scheme for OFDM communication waveforms, while Johnston \emph{et al.} \cite{9729203} focused on jointly designing transmission waveforms and reception filters in multiple-input multiple-output (MIMO-OFDM) systems to meet specific requirements like bit error rate (BER) and directional radiated power.
    \item Radar-centric design (RCD): RCD prioritizes sensing performance, incorporating communication data into radar waveforms. A common method is using radar waveforms as carriers for modulating communication symbols. For example, \cite{8008137,9083884} employed frequency modulated continuous wave (FMCW) and chirp waveforms as carriers for communication information. Another approach, seen in \cite{7060416,7347464}, involves selecting radar waveforms from a set to form a communication symbol set. Li \cite{10279030,10283671} suggested using a set of radar waveform filters as a communication codebook.
    \item Joint design (JD): JD aims to find a flexible balance between communications and radar through optimization-based signal designs \cite{8386661,2021JointTW,9104378,9534484,9850347,8683591,9652071}. The work in \cite{8386661} introduced a compromise factor for minimizing radar waveform mismatch and reducing multi-user interference in communications. This concept has influenced subsequent research, such as \cite{2021JointTW,9104378}. Liu \emph{et al.} in \cite{9534484} proposed a constructive interference (CI) symbol-level precoding to satisfy both radar beampattern mismatch and communication quality of service (QoS) requirements. To simplify the implementation, \cite{9850347} introduced block-level ISAC precoding for finite alphabet inputs. Meanwhile, \cite{8683591} considered ISAC signal design with hybrid beamforming and sub-connected structures to reduce hardware complexity and cost. Guo \emph{et al.} \cite{10.1145/3556562.3558569} proposed constructing an ISAC beamformer candidate set for scenarios with limited RF chains to enhance spectral efficiency. Lastly, Liu \emph{et al.} in \cite{9652071} explored minimizing the estimation Cramér-Rao bound (CRB) for sensing while meeting communication signal-to-interference-and-noise ratio (SINR) requirements.
\end{itemize}

It's important to note that all the aforementioned studies introduce certain inherent constraints in comparison to the pure ISAC signal design problem (\textbf{P2}). For instance, research aiming to maximize spectral efficiency under Shannon's theorem inherently assumes a Gaussian input constraint \cite{8683591}. Similarly, studies focused on minimizing BER or managing symbol-level interferences typically operate under the assumption that transmitted or shaped data symbols are of $M$-QAM or $M$-PSK formats \cite{8386661,2021JointTW,9104378,9534484,9850347}. These additional constraints can potentially compromise the effectiveness of ISAC signals. As a result, the signals or waveforms derived under these conditions may be sub-optimal compared to those that would be obtained by directly solving the pure ISAC signal design problem (\textbf{P2})

\subsection{Contributions}
As introduced earlier, the design of communication signals inherently resembles a sphere packing problem. The integration of sensing functionality adds an additional constraint to this problem, forming (\textbf{P2}), a concept not fully explored in previous studies. This paper seeks to tackle this fundamental problem without imposing any extraneous constraints. The primary contributions of this work are outlined as follows:
\begin{itemize}
    \item We present a unified approach to formulating the ISAC signal design problem across time, frequency, and space domains. This culminates in a modified sphere packing problem aimed at maximizing the minimum Euclidean distance (MMED) between high-dimensional signals, considering both sensing waveform constraints and power limitations. 
    \item  Through meticulous formula manipulation, we transform the generalized ISAC signal design problem into a large-scale quadratic constrained quadratic programming (QCQP) problem. We introduce an augmented Lagrangian and dual ascent (ALDA) algorithm to address this challenge, with a detailed analysis of the computational complexity.
    \item To mitigate the computational complexity, we propose a bit-dimension-power splitting (BDPS) method. This approach divides the objective function and constraints, converting the overarching QCQP problem into a series of smaller-scale QCQP sub-problems. These can be solved in parallel, significantly reducing computational demands.
    \item Extensive simulations validate the efficacy of our algorithms. Through thorough comparisons with existing works, we demonstrate the superiority of our ISAC signal designs. Our results indicate that our proposed designs achieve a lower SER and higher MI while maintaining robust sensing performance. Moreover, the BDPS algorithm effectively lowers complexity with minimal performance degradation.
\end{itemize}
\subsection{Organization}
The remainder of this paper is organized as follows.
 In Section II, we delve into the formulation of ISAC signal design problems across time, frequency, and spatial domains. We then unify these problems to establish a comprehensive framework. Section III is dedicated to the transformation of the unified ISAC signal design problem. In Section IV, we introduce the ALDA solution, a method we've developed to solve the transformed ISAC signal design problem. In Section V, we discuss the BDPS solution. This method is proposed as a strategy to reduce the computational complexity of the ISAC signal design problem. In Section VI, we present the results of our simulations. This section includes a thorough discussion of the outcomes, providing insights into the effectiveness and efficiency of the proposed methods. Finally, Section VII concludes the paper.
\subsection{Notations}
The following notations are used in this paper. Boldface lower-case letter $\mathbf{s}$, upper-case letter $\mathbf{S}$, and non-bolded letter $s$ or $S$ indicate column
vectors, matrices, and scalar, respectively. $\mathcal{S}$ stands for a set and $|\mathcal{S}|$ is the size of the set. $(\cdot)^T$ and $(\cdot)^H$ denote the
transpose and the transpose-conjugate operations, respectively. $\Im\{\cdot\}$ and $\Re\{\cdot\}$ denote the imaginary and real part of a complex number. $\mathbb{R}$ and $\mathbb{C}$ stand for the real-number space and complex-number space, respectively. $\|\cdot\|_2$ denotes the $l_2$ norm. $\otimes$ and $\odot$ represent the Kronecker product and Hadamard product of two matrices or vectors, respectively. $\mathbf{I}_N$ represents a $N\times N$ identity matrix and $\mathbf{1}_N$ represents all ones vector of length $N$. $[x]^+$ stand for the function $\max\{x,0\}$. Finally, $\diag(\mathbf{s})$ stands for a diagonal matrix with
diagonal elements being $\mathbf{s}$.

\section{Problem Description and Generalization}
\begin{figure}[!t]
       \centering
        \subfigure[Time signal]{{\label{fig1a}}\includegraphics[width=0.9\linewidth]{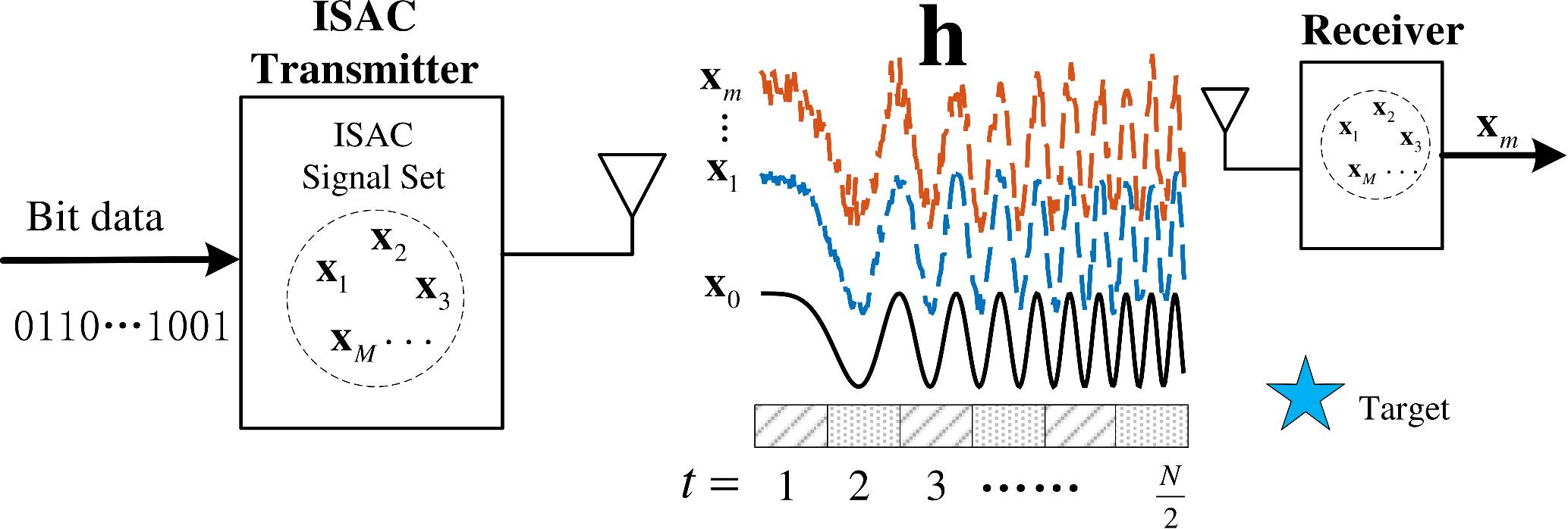}}
         \subfigure[Frequency signal]{{\label{fig1b}}\includegraphics[width=0.9\linewidth]{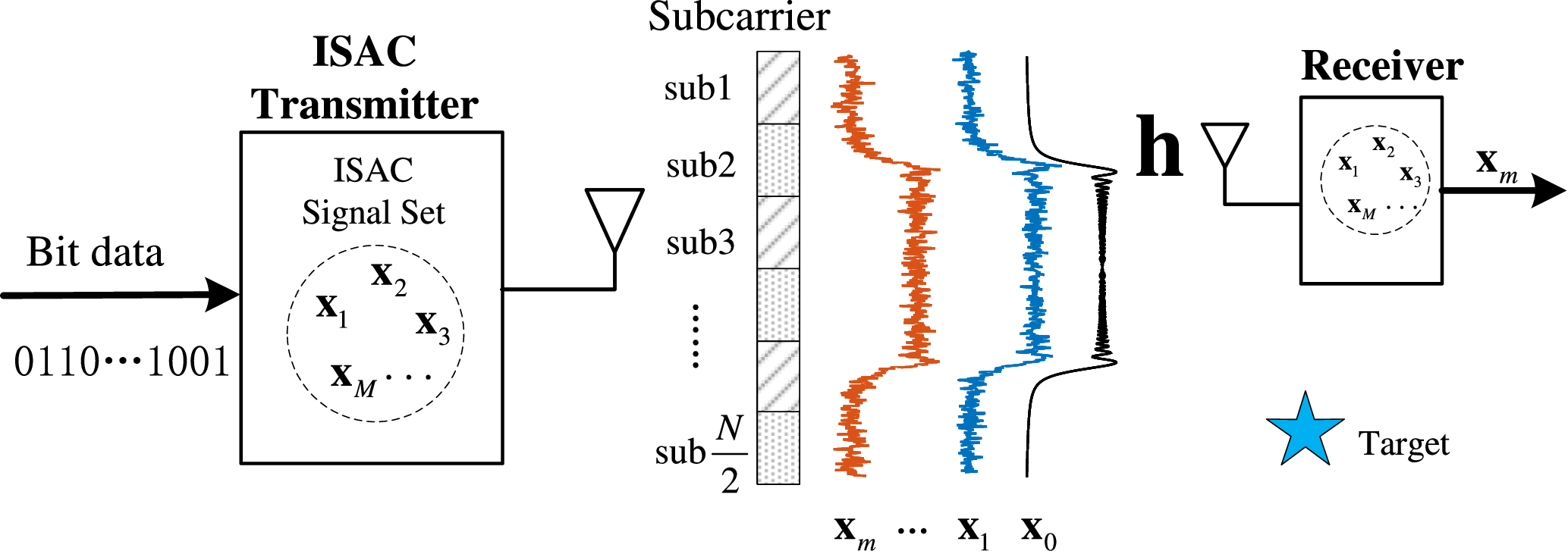}}
        \subfigure[Spatial signal]{{\label{fig1c}}\includegraphics[width=0.9\linewidth]{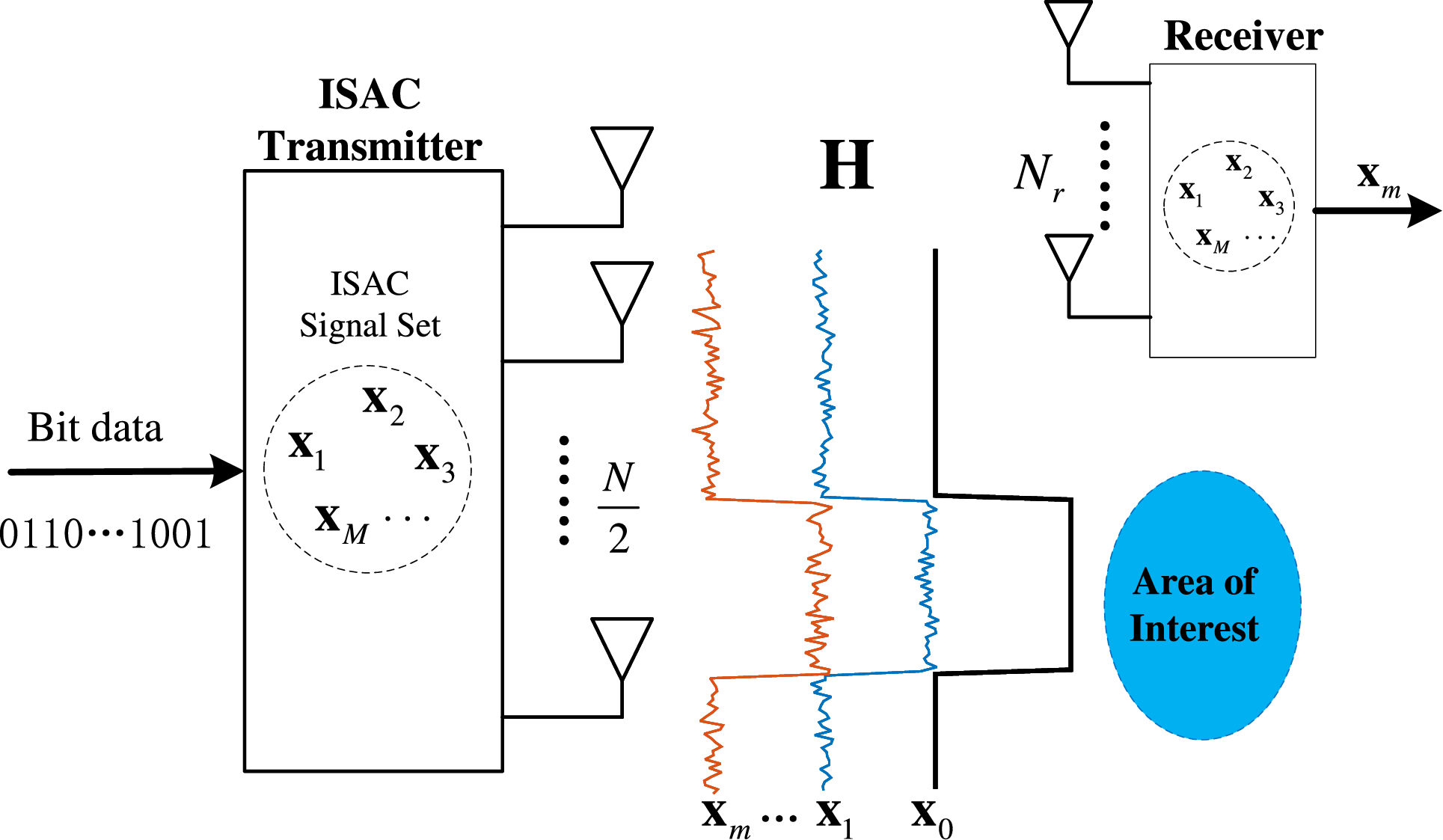}}
        \caption{Schematic of the proposed unified ISAC signaling design.}
         \label{fig1}
\end{figure}

In this paper, we proceed under the assumption that the reference sensing signal, denoted as $\mathbf{x}_0$, has already been identified. The goal for reference ISAC signals is to ensure they are closely aligned with $\mathbf{x}_0$ to maintain satisfactory sensing performance. { According to the definition in \(\textbf{P2}\), the core challenge lies in identifying a set \(\mathcal{X}\) of real-valued signals containing \(M\) elements, where each signal has a dimension of \(N\). Each \(N\)-dimensional real-valued signal corresponds to \(N/2\)-dimensional complex signal, which is synthesized from the in-phased and quadrature components in communication systems. Each \(N/2\)-dimensional complex symbol needs to be transmitted over \(N/2\) time, frequency, or spatial resource blocks, as illustrated in Fig. \ref{fig1}.}

To comprehensively address the ISAC signal design, we consider the following three scenarios without loss of generality:
\begin{itemize}
    \item The first scenario involves designing $N/2$-dimensional complex ISAC signals without any CSI. This scenario directly tackles the original problem (\textbf{P2}).
    \item The second scenario focuses on designing $N/2$-dimensional complex ISAC signals with the knowledge of CSI at the transmitter (CSIT). In this case, the $N/2$-dimensional signal undergoes independent fading, free from mutual interference. This could occur, for instance, when $N/2$-dimensional complex symbols within a signal are transmitted over orthogonal resources, such as different time slots or orthogonal sub-carriers. The optimization problem in this scenario can be described as a modified version of (\textbf{P2}), taking into account the CSIT. The modified problem formulation could be represented as follows:
\begin{align}
(\textbf{P3}):~\mathrm{Given}: &~M,N,P,\mathbf{A},\mathbf{x}_0,\epsilon\notag\\
\mathrm{Find}:&~\mathcal{X}=\{\mathbf{x}_1,\mathbf{x}_2,\cdots,\mathbf{x}_M\}\notag\\
\mathrm{Maximize}:&~\min_{\forall k,l,k\neq l} ||\mathbf{A}(\mathbf{x}_k-\mathbf{x}_l)||_2\\
\mathrm{Subject~to}:&~\frac{1}{M}\sum_{k=1}^{M}||\mathbf{x}_k||_2^2\leq P,\notag\\
&~||\mathbf{x}_k-\mathbf{x}_0||_2\leq \epsilon, \forall k\notag,
\end{align}
where $\mathbf{A}=\diag([\Re(\mathbf{h})^T,\Im(\mathbf{h})^T])\in\mathbb{R}^{N\times N}$ and $\mathbf{h}\in \mathbb{C}^{N/2}$ denotes the channel vector of $N/2$ independent fading channels without mutual interference.
\item The third scenario corresponds to designing $N/2$-dimensional complex ISAC signals with CSIT, where the signal experiences channels with crosstalk. An example is the transmission of $N/2$-dimensional complex symbols through MIMO channels with inter-channel interferences (ICI). Let $N_r$ represent the degrees of freedom of the received signal, and $\mathbf{H}$ be a matrix in $\mathbb{C}^{N_r\times N/2}$, characterizing the channel. The optimization problem in this scenario involves a further modification of (\textbf{P2}), taking into account both CSIT and the unique channel characteristics due to crosstalk. It can be expressed as
\begin{align}
(\textbf{P4}):~\mathrm{Given}: &~M,N,P,\mathbf{B},\mathbf{x}_0,\epsilon\notag\\
\mathrm{Find}:&~\mathcal{X}=\{\mathbf{x}_1,\mathbf{x}_2,\cdots,\mathbf{x}_M\}\notag\\
\mathrm{Maximize}:&~\min_{\forall k,l,k\neq l} ||\mathbf{B}(\mathbf{x}_k-\mathbf{x}_l)||_2\\
\mathrm{Subject~to}:&~\frac{1}{M}\sum_{k=1}^{M}||\mathbf{x}_k||_2^2\leq P,\notag\\
&~||\mathbf{x}_k-\mathbf{x}_0||_2\leq \epsilon, \forall k\notag,
\end{align}
where 
\begin{equation}
\mathbf{B}= \left[ \begin{array}{ccc}
\Re{(\mathbf{H})} & -\Im{(\mathbf{H})} \\
\Im{(\mathbf{H})} & \Re{(\mathbf{H})}\end{array} \right]\in\mathbb{R}^{2N_r\times N}.
\end{equation}

We make the singular value composition (SVD) of $\mathbf{B}$ as $\mathbf{B}=\mathbf{U}\pmb{\Lambda}\mathbf{V}^T$, where $\mathbf{U}$ is a $2N_r\times 2N_r$ complex unitary matrix,  $\pmb{\Lambda}$ is a $2N_r\times N$ rectangular diagonal matrix with non-negative real numbers on the diagonal, $\mathbf{V}$ is a $N\times N$ real unitary matrix. By introducing $\mathbf{s}_i\in\mathbb{R}^{N}$ and letting $\mathbf{x}_i=\mathbf{V}\mathbf{s}_i$, $i=1,2,\cdots,M$, we can express the term $||\mathbf{B}(\mathbf{x}_k-\mathbf{x}_l)||_2$ in (\textbf{P4}) as
\begin{equation}
||\mathbf{B}(\mathbf{x}_k-\mathbf{x}_l)||_2=||\pmb{\Lambda}(\mathbf{s}_k-\mathbf{s}_l)||_2.
\end{equation}
and the constraint of (\textbf{P4})  as
\begin{equation}
\frac{1}{M}\sum_{k=1}^{M}||\mathbf{s}_k||_2^2\leq P,
\end{equation}
and 
\begin{equation}
||\mathbf{s}_k-\mathbf{s}_0||_2\leq \epsilon.
\end{equation}
Then, we can reformulate $(\textbf{P4})$ as
\begin{align}
(\textbf{P5}):~\mathrm{Given}: &~M,N,P,\pmb{\Lambda},\mathbf{s}_0,\epsilon\notag\\
\mathrm{Find}:&~\mathcal{S}=\{\mathbf{s}_1,\mathbf{s}_2,\cdots,\mathbf{s}_M\}\notag\\
\mathrm{Maximize}:&~\min_{k,l\in [1,M],k\neq l} ||\pmb{\Lambda}(\mathbf{s}_k-\mathbf{s}_l)||_2\\
\mathrm{Subject~to}:&~\frac{1}{M}\sum_{k=1}^{M}||\mathbf{s}_k||_2^2\leq P,\notag\\
&~||\mathbf{s}_k-\mathbf{s}_0||_2\leq \epsilon, \forall k\notag.
\end{align}
Solving (\textbf{P5}) is equivalently solving (\textbf{P4}) as $\mathbf{x}_i=\mathbf{V}\mathbf{s}_i$, $i=1,2,\cdots,M$.
\end{itemize}
Upon careful examination of the formulations of (\textbf{P2}), (\textbf{P3}), and (\textbf{P5}), it becomes apparent that these ISAC signal shaping problems share notable similarities. Notably, (\textbf{P2}) can be viewed as a specific instance of (\textbf{P3}) when we set $\mathbf{A}$ to be the identity matrix $\mathbf{I}_{N}$. Similarly, (\textbf{P3}) can be considered a particular case of (\textbf{P5}) under the condition that $N_r$ is equal to $N/2$. This hierarchical relationship among the problems suggests that a solution to the more general (\textbf{P5}) would inherently encompass solutions to the more specific scenarios presented in (\textbf{P2}) and (\textbf{P3}). Therefore, our subsequent analysis will primarily focus on addressing and solving the general problem (\textbf{P5}). Through this approach, we aim to develop a comprehensive solution strategy that is applicable across the varying contexts of (\textbf{P2}) and (\textbf{P3}).

\section{Problem Transformation}
Problem (\textbf{P5}) is a modified sphere packing problem with weighted distance maximization and additional sensing constraints. We aim to propose a generalized solution to this problem. To fulfill this goal, we transform the original problem into a simplified version. To do so, we introduce several auxiliary matrices and vectors. In detail,  we stack $M$ $\pmb{\Lambda}$ matrices into
\begin{equation}
\mathbf{G}=[\overbrace{\pmb{\Lambda},\pmb{\Lambda},\cdots,\pmb{\Lambda}}^{M}]\in\mathbb{R}^{2N_r\times MN},
\end{equation}
and all variable vectors into
\begin{equation}
\mathbf{z}=[\mathbf{s}_1^T,\mathbf{s}_2^T,\cdots,\mathbf{s}_M^T]^T\in\mathbb{R}^{MN}.
\end{equation}
By introducing
\begin{equation}
\mathbf{D}_{\mathbf{z}}=\diag(\mathbf{z})\in\mathbb{R}^{MN\times MN},
\end{equation}
and 
\begin{equation}
\mathbf{e}_k=\mathbf{g}_k\otimes \mathbf{1}_{N}\in\mathbb{R}^{MN},
\end{equation}
where $\mathbf{g}_k\in \mathbb{R}^{M}$ is a standard basis vector with all zeros except a one at the $k$th position, we can express the term $||\pmb{\Lambda}(\mathbf{s}_k-\mathbf{s}_l)||_2$ as
\begin{equation}
||\pmb{\Lambda}(\mathbf{s}_k-\mathbf{s}_l)||_2=||\mathbf{G}\mathbf{D}_{\mathbf{z}}(\mathbf{e}_k-\mathbf{e}_l)||_2.
\end{equation}
Taking the square of the term $||\pmb{\Lambda}(\mathbf{s}_k-\mathbf{s}_l)||_2$, we have
\begin{equation}\label{eq15}
\begin{split}
||\pmb{\Lambda}(\mathbf{s}_k-\mathbf{s}_l)||_2^2&=(\mathbf{e}_k-\mathbf{e}_l)^T \mathbf{D}_{\mathbf{z}}^T\mathbf{G}^T\mathbf{G}\mathbf{D}_{\mathbf{z}}(\mathbf{e}_k-\mathbf{e}_l)\\
&=\tr (\mathbf{D}_{\mathbf{z}}^T \mathbf{R}_{\mathbf{G}}\mathbf{D}_{\mathbf{z}} \mathbf{W}_{kl})
\end{split}
\end{equation}
where $\mathbf{R}_G=\mathbf{G}^T\mathbf{G}$ and $\mathbf{W}_{kl}=(\mathbf{e}_k-\mathbf{e}_l)(\mathbf{e}_k-\mathbf{e}_l)^T$. According to $\tr(\mathbf{D}_\mathbf{u}\tilde{\mathbf{A}}{\mathbf{D}}_{\mathbf{v}}\tilde{\mathbf{B}}^T)=\mathbf{u}^T(\tilde{\mathbf{A}}\odot\tilde{\mathbf{B}})\mathbf{v}$ \cite{zhang2017matrix},  with $\mathbf{D}_{\mathbf{u}}=\diag (\mathbf{u})$ and $\mathbf{D}_{\mathbf{v}}=\diag (\mathbf{v})$, we can rewrite (\ref{eq15}) as
\begin{equation}
||\pmb{\Lambda}(\mathbf{s}_k-\mathbf{s}_l)||_2^2=\mathbf{z}^T\mathbf{Q}_{kl}\mathbf{z},
\end{equation}
where $\mathbf{Q}_{kl}=\mathbf{R}_{G}\odot \mathbf{W}_{kl}^T\in\mathbb{R}^{MN\times MN}$.

The first constraint of (\textbf{P5}) can be written as
\begin{equation}
\mathbf{z}^T\mathbf{z}\leq MP.
\end{equation}
By introducing $\mathbf{E}_k=\diag(\mathbf{e}_k)$, we can rewrite the sensing constraints in (\textbf{P5}) as 
\begin{equation}
\begin{split}
||\mathbf{s}_k-\mathbf{s}_0||_2^2&=||\mathbf{E}_k(\mathbf{z}-\mathbf{z}_0)||_2^2\\
&=(\mathbf{z}-\mathbf{z}_0)^T\mathbf{E}_k(\mathbf{z}-\mathbf{z}_0)\\
&\leq \epsilon^2, \forall k,
\end{split}
\end{equation}
where 
\begin{equation}
\mathbf{z}_0=[\overbrace{\mathbf{s}_0^T,\mathbf{s}_0^T,\cdots,\mathbf{s}_0^T}^{M}]^T\in\mathbb{R}^{MN},
\end{equation}
and the equation holds due to the fact $\mathbf{E}_k=\mathbf{E}_k^T\mathbf{E}_k$.

By above skillful formula manipulations, we can derive a simplified but equivalent version of (\textbf{P5}) as
{
\begin{align}
(\textbf{P6}):~\mathrm{Given}: &~M,N,P,\pmb{\Lambda},\mathbf{z}_0,\epsilon\notag\\
\mathrm{Find}:&~\mathbf{z}=[\mathbf{s}_1^T,\mathbf{s}_2^T,\cdots,\mathbf{s}_M^T]^T\in\mathbb{R}^{MN}\notag\\
\mathrm{Maximize}:&~\min_{\forall k,l,k\neq l} \mathbf{z}^T\mathbf{Q}_{kl}\mathbf{z}\\
\mathrm{Subject~to}:&~\mathbf{z}^T\mathbf{z}\leq MP,\notag\\
&~(\mathbf{z}-\mathbf{z}_0)^T\mathbf{E}_k(\mathbf{z}-\mathbf{z}_0)\leq \epsilon^2, \forall k\notag.
\end{align}
}
By introducing an auxiliary variable $t$, we have the equivalent epigraph form \cite{boyd2004convex} of (\textbf{P6}) as
{
\begin{align}
(\textbf{P7}):~\mathrm{Given}: &~M,N,P,\pmb{\Lambda},\mathbf{z}_0,\epsilon\notag\\
\mathrm{Find}:&~\mathbf{z}=[\mathbf{s}_1^T,\mathbf{s}_2^T,\cdots,\mathbf{s}_M^T]^T\in\mathbb{R}^{MN}\notag\\
\mathrm{Maximize}:&~t\\
\mathrm{Subject~to}:&~\mathbf{z}^T\mathbf{z}\leq MP,\notag\\
&~\mathbf{z}^T\mathbf{Q}_{kl}\mathbf{z}\geq t, \forall k,l, k\neq l,\notag\\
&~(\mathbf{z}-\mathbf{z}_0)^T\mathbf{E}_k(\mathbf{z}-\mathbf{z}_0)\leq \epsilon^2, \forall k\notag.
\end{align}
}
By introducing a target minimum distance $d$, we can introduce a new problem closely related to (\textbf{P7}) as
{
\begin{align}
(\textbf{P8}):~\mathrm{Given}: &~M,N,P,\pmb{\Lambda},\mathbf{z}_0,\epsilon,d\notag\\
\mathrm{Find}:&~\mathbf{z}=[\mathbf{s}_1^T,\mathbf{s}_2^T,\cdots,\mathbf{s}_M^T]^T\in\mathbb{R}^{MN}\notag\\
\mathrm{Minimize}:&~\mathbf{z}^T\mathbf{z}\\
\mathrm{Subject~to}:&~\mathbf{z}^T\mathbf{Q}_{kl}\mathbf{z}\geq d, \forall k,l, k\neq l,\notag\\
&~(\mathbf{z}-\mathbf{z}_0)^T\mathbf{E}_k(\mathbf{z}-\mathbf{z}_0)\leq \epsilon^2, \forall k\notag.
\end{align}
}
Given $\mathbf{z}^*$ as the solution to (\textbf{P8}), and letting ${P}^*$ represent the average power, i.e., ${P}^*=\frac{1}{M}(\mathbf{z}^*)^T\mathbf{z}^*$, there are three cases.
\begin{itemize}
    \item If ${P}^*$ is  equal to ${P}$, the solution to (\textbf{P8}) is exactly the solution to (\textbf{P7}) and we have $t=d$.
     \item If ${P}^*$ is smaller than ${P}$, the original (\textbf{P7}) is feasible with an objective value $t>d$. However, $z^*$ is not the optimal solution to (\textbf{P7}), because we can still increase the communication and sensing performance with an additional power budget (i.e., $P-P^*$).
     \item If ${P}^*$ is greater than ${P}$, the original (\textbf{P7}) is infeasible.
\end{itemize}

From our analysis, we identify an optimal $d^*$ using one-dimensional optimization methods, such as the bisection method, to ensure that $P^*=P$. Consequently, solving problem (\textbf{P8}) with this $d^*$ becomes equivalent to solving (\textbf{P7}). The crux of the issue lies in efficiently solving (\textbf{P8}), a QCQP problem with $MN$ variables and $\frac{M(M+1)}{2}$ constraints.

The QCQP challenge in (\textbf{P8}) can be approached using various optimization strategies, such as semidefinite relaxation (SDR) techniques \cite{5447068}. SDR approximates a solution by transforming the original problem into a semi-definite programming (SDP) problem. Specifically for (\textbf{P8}), this involves setting $\mathbf{C} = \mathbf{z}\mathbf{z}^H$ and relaxing the rank-1 constraint of matrix $\mathbf{C}$. A feasible solution to the original problem is then derived by reinstating the rank-1 constraint, using methods like Gaussian randomization. However, the direct application of SDR is hindered by the sheer volume of quadratic constraints and optimization variables, such as $1024$ variables and $136$ constraints for $N/2=32, M=16$. The extensive number of quadratic constraints significantly reduces the likelihood of obtaining rank-1 solutions.

To overcome these challenges, this paper introduces an augmented Lagrangian and dual ascent (ALDA)-based algorithm. This algorithm is specifically designed to address the complexities associated with the large number of constraints and variables in (\textbf{P8}).

\section{Signal Set Design Using ALDA Algorithm}
In this section, we present the development of the ALDA algorithm, which employs augmented Lagrangian and dual ascent techniques to address (\textbf{P8}). Initially, we provide a concise overview of the fundamental principles underlying the dual ascent method. Following this, we delve into a detailed exposition of the ALDA algorithm.

\subsection{Dual Ascent}
Dual ascent is a strategy for solving convex optimization problems\cite{Boyd2011DistributedOA}. 
 { Let's consider a general constrained convex optimization problem with equality constraints and non-equality constraints as follows:

\begin{equation}
\begin{split}
    \label{4-1}
    \min_{\mathbf{x}\in \mathbb{R}} &~f(\mathbf{x})\\
    \mathrm{subject~to:}&~c_i(\mathbf{x})=0,~ i \in \mathcal{I},\\
    &~c_j(\mathbf{x})\leq 0,~j \in \mathcal{M},
 \end{split}
\end{equation}
where \(\mathcal{I}\) and \(\mathcal{M}\) are the index sets for equality and inequality constraints, respectively.

We can construct the Lagrangian function
\begin{equation}
\begin{split}
    \label{4-2}
    L(\mathbf{x},\pmb{\lambda},\mathbf{v})=&f(\mathbf{x})+\sum_{i\in \mathcal{I}}\lambda_i c_i(\mathbf{x})+\sum_{j\in \mathcal{M}}v_j c_j(\mathbf{x}),
\end{split}
\end{equation}
where $\pmb{\lambda}\triangleq\{\lambda_i\},\mathbf{v}\triangleq\{v_j\}$ are the dual variables. }In the feasible domain of (\ref{4-1}) $\mathcal{D}=\{\mathbf{x}|c_i(\mathbf{x})=0, i \in \mathcal{I},c_j(\mathbf{x})\leq0,j \in \mathcal{M}\}$, the original function $f(\mathbf{x})$ is the upper bound of Lagrangian function (\ref{4-2}), i.e.,
\begin{equation}
    \label{4-3}
    \max_{\pmb{\lambda},\mathbf{v}}L(\mathbf{x},\pmb{\lambda},\mathbf{v})=f(\mathbf{x}), ~\mathbf{x}\in \mathcal{D}.
\end{equation}
Therefore, the optimization of the original problem can be equivalently written as
\begin{equation}
\label{4-4}
    \min_\mathbf{x} f(\mathbf{x})= \min_\mathbf{x} \max_{\pmb{\lambda},\mathbf{v}}L(\mathbf{x},\pmb{\lambda},\mathbf{v}),~\mathbf{x}\in \mathcal{D}.
\end{equation}
The dual problem
with respect to the original problem in (\ref{4-1}) can be expressed as
\begin{equation}
    \label{4-5}
    \max_{\pmb{\lambda},\mathbf{v}}\min_\mathbf{x}L(\mathbf{x},\pmb{\lambda},\mathbf{v}).
\end{equation}
when the convex optimization problem satisfies strong duality, the dual problem has the same optimal solution as the original problem. Based on this, one can update $\mathbf{x}$ and $\pmb{\lambda},\mathbf{v}$ alternatively. {Specifically, in the $j$th step, one can update $\mathbf{x}$ to be
\begin{equation}
    \label{4-6}
    \mathbf{x}^{[j+1]}=\arg \min_{\mathbf{x}} L(\mathbf{x},\pmb{\lambda}^{[j]},\mathbf{v}^{[j]})
\end{equation}}{
For the unconstrained \( \mathbf{x} \)-minimization problem, solutions can be obtained using optimization methods such as gradient descent and quasi-Newton methods \cite{boyd2004convex,8267071,8186925}. }Afterwards, one can update $\pmb{\lambda},\mathbf{v}$ to be
\begin{equation}
\pmb{\lambda}^{[j+1]}=\arg \max_{\pmb{\lambda}} L(\mathbf{x}^{[j+1]},\pmb{\lambda},\mathbf{v}^{[j]})
\end{equation}
and
\begin{equation}
    \mathbf{v}^{[j+1]}=\arg \max_{\mathbf{v}} L(\mathbf{x}^{[j+1]},\pmb{\lambda}^{[j+1]},\mathbf{v}).
\end{equation}
For convenience, the gradient ascent method is often used in updating the dual variables \footnote{Gradient ascent is just the opposite of gradient descent. Gradient ascent adjusts the parameters in the direction of the gradient to maximize some objective function \cite{9427806}.}, i.e.,
\begin{equation}
    \label{4-8a}\pmb{\lambda}^{[j+1]}=\pmb{\lambda}^{[j]}+\eta^j\Delta_{\pmb{\lambda}} L(\mathbf{x}^{[j+1]},\pmb{\lambda},\mathbf{v}^{[j]})
\end{equation}
and
\begin{equation}    
\label{4-8b}\mathbf{v}^{[j+1]}=\mathbf{v}^{[j]}+\eta^j\Delta_\mathbf{v}L(\mathbf{x}^{[j+1]},\pmb{\lambda}^{[j+1]},\mathbf{v})
\end{equation}
where $\eta^j$ is the step size in the $j$-th iteration, $\Delta_{\pmb{\lambda}}L(\mathbf{x}^{[j+1]},\pmb{\lambda},\mathbf{v}^{[j]})$ denotes the derivative with respect to $\pmb{\lambda}$, and $\Delta_{\mathbf{v}}L(\mathbf{x}^{[j+1]},\pmb{\lambda}^{[j+1]},\mathbf{v})$ denotes the derivative with respect to $\mathbf{v}$. By iteratively updating these parameters using equations (\ref{4-6}), (\ref{4-8a}) and (\ref{4-8b}), the solutions to the dual problem and the primal problem can be obtained.

\subsection{ALDA Algorithm for Solving (\textbf{P8})}
For the inequality constraint in (\textbf{P8}), we equivalently transform it into the following optimization problem by introducing slack variables,
{
\begin{align}
(\textbf{P9}):~\mathrm{Given}: &~M,N,P,\pmb{\Lambda},\mathbf{z}_0,\epsilon,d\notag\\
\mathrm{Find}:&~\mathbf{z}=[\mathbf{s}_1^T,\mathbf{s}_2^T,\cdots,\mathbf{s}_M^T]^T\in\mathbb{R}^{MN}\notag\\&~\mathbf{r}\triangleq\{r_{k,l}\}, \forall k,l\in\{1,2,\cdots,M\},k\neq l\notag\\&~\mathbf{t}\triangleq \{t_k\}, \forall  k\in\{1,2,\cdots,M\}\notag\\
\mathrm{Minimize}:&~\mathbf{z}^T\mathbf{z}\\
\mathrm{Subject~to}:&~d-\mathbf{z}^T\mathbf{Q}_{kl}\mathbf{z}+r_{k,l}=0 , \forall k,l, k\neq l,\notag\\
&~\|\mathbf{E}_k(\mathbf{z}-\mathbf{z}_0)\|_2^2-\epsilon^2+t_k= 0, \forall k\notag,\\
&r_{k,l}\geq0,t_k\geq0\notag,
\end{align}
}
where $\mathbf{r}\triangleq\{r_{k,l}\},\mathbf{t}\triangleq \{t_k\}$ are the introduced slack variables.
Then, we can write the augmented Lagrangian function for (\textbf{P9}) as
\begin{equation}
\begin{split}
    \label{eq34}
&L(\mathbf{z},\mathbf{r},\mathbf{t},\pmb{\lambda},\mathbf{v},\mu)\\&~=\mathbf{z}^T\mathbf{z}-\sum_{l=1}^{M}\sum_{k=1,k\neq l}^{M} \lambda_{l,k}\left(\mathbf{z}^T\mathbf{Q}_{kl}\mathbf{z}-r_{k,l}-d \right)\\
    &~~~~~+\sum_{k=1}^{M}v_k \left(\|\mathbf{E}_k(\mathbf{z}-\mathbf{z}_0)\|_2^2-\epsilon^2+t_k\right)\\
    &~~~~~+\frac{\mu}{2}\left[\sum_{l=1}^{M}\sum_{k=1,k\neq l}^{M} \left(\mathbf{z}^T\mathbf{Q}_{kl}\mathbf{z}-r_{k,l}-d \right)^2\right.\\
    &~~~~~~~~~~~~~+\left.\sum_{k=1}^{M} \left(\|\mathbf{E}_k(\mathbf{z}-\mathbf{z}_0)\|_2^2+t_k-\epsilon^2\right)^2\right],
\end{split}
\end{equation}
where $\pmb{\lambda}\triangleq\{\lambda_{l,k}\},\mathbf{v}\triangleq\{v_k\}$ are the dual variables,  {the penalty terms $\sum_{l=1}^{M}\sum_{k=1,k\neq l}^{M} \left(\mathbf{z}^T\mathbf{Q}{kl}\mathbf{z}-r{k,l}-d \right)^2$ and $\sum_{k=1}^{M} \left(|\mathbf{E}_k(\mathbf{z}-\mathbf{z}0)|2^2+t_k-\epsilon^2\right)^2$ correspond to the constraints $d-\mathbf{z}^T\mathbf{Q}{kl}\mathbf{z}+r{k,l}=0$ and $|\mathbf{E}_k(\mathbf{z}-\mathbf{z}_0)|_2^2-\epsilon^2+t_k=0$, respectively. These quadratic terms, controlled by the penalty parameter $\mu>0$, enable the optimization process to effectively approximate the constraints, thereby ensuring strict satisfaction of these constraints in the final solution.}

According to the principle of the dual ascent method, we will update the parameters alternately. In the $j$-th iteration step, given the dual variables $\pmb{\lambda}^{[j]},\mathbf{v}^{[j]}$ and the penalty parameter $\mu^{[j]}$, we can update $\mathbf{z}$, $\mathbf{r}$, and $\mathbf{t}$ as follows. 
\subsubsection{Update $\mathbf{z}$, $\mathbf{r}$, and $\mathbf{t}$}
According to (\ref{4-6}), we need to solve the following problem to find $\mathbf{z}^{[j+1]}$, $\mathbf{r}^{[j+1]}$ and $\mathbf{t}^{[j+1]}$:
\begin{equation}
\begin{split}
\label{4-11}
        \min_{\mathbf{z},\mathbf{r},\mathbf{t}} &~L(\mathbf{z},\mathbf{r},\mathbf{t},\pmb{\lambda}^{[j]},\mathbf{v}^{[j]},\mu^{[j]})\\
        \mathrm{subject~to:}&~\mathbf{r}\succeq\mathbf{0},\mathbf{t}\succeq\mathbf{0}.
\end{split}
\end{equation}
The dominant terms in (\ref{eq34}) to update slack variables $\mathbf{r}$ and $\mathbf{t}$ can be written as
\begin{equation}
\begin{split}
    &\min_{\mathbf{r}\succeq \mathbf{0}} ~\sum_{l=1}^{M}\sum_{k=1,k\neq l}^{M} -\lambda_{l,k}^{[j]}\left(\mathbf{z}^T\mathbf{Q}_{kl}\mathbf{z}-r_{k,l}-d \right)\\
    &~~~~~+\frac{\mu^{[j]}}{2}\sum_{l=1}^{M}\sum_{k=1,k\neq l}^{M} \left(\mathbf{z}^T\mathbf{Q}_{kl}\mathbf{z}-r_{k,l}-d \right)^2,
    \end{split}
\end{equation}
and
\begin{equation}
\begin{split}
   &\min_{\mathbf{t}\succeq \mathbf{0}}~ \sum_{k=1}^{M}v_k^{[j]} \left(\|\mathbf{E}_k(\mathbf{z}-\mathbf{z}_0)\|_2^2-\epsilon^2+t_k\right)\\
    &~~~~~+\frac{\mu^{[j]}}{2}\sum_{k=1}^{M} \left(\|\mathbf{E}_k(\mathbf{z}-\mathbf{z}_0)\|_2^2+t_k-\epsilon^2\right)^2.
    \end{split}
    \end{equation}

Taking the derivatives of the objective functions with respect to $r_{l,k},v_k$, we can derive the zero point to obtain the updates as
\begin{equation}\label{eq38}
    r_{l,k}^{[j+1]}=\left[\mathbf{z}^T\mathbf{Q}_{kl}\mathbf{z}-d-\frac{\lambda_{l,k}^{[j]}}{\mu^{[j]}}\right]^+,
\end{equation}
and
\begin{equation}\label{eq39}
    t_k^{[j+1]}=\left[-\|\mathbf{E}_k(\mathbf{z}-\mathbf{z}_0)\|_2^2+\epsilon^2-\frac{v_k^{[j]}}{\mu^{[j]}}\right]^+.
\end{equation}

\begin{figure*}
\normalsize
\setcounter{MYtempeqncnt}{\value{equation}}
\setcounter{equation}{39}
    
    \begin{align}
    \label{4-14}
            &-\lambda_{l,k}^{[j]}\left(\mathbf{z}^T\mathbf{Q}_{kl}\mathbf{z}-r_{k,l}-d \right)+\frac{\mu^{[j]}}{2}\left(\mathbf{z}^T\mathbf{Q}_{kl}\mathbf{z}-r_{k,l}-d \right)^2\notag\\
            &~~=\left\{
            \begin{array}{l}
            -\lambda_{l,k}^{[j]}\left(\mathbf{z}^T\mathbf{Q}_{kl}\mathbf{z}-d \right)+\frac{\mu^{[j]}}{2}\left(\mathbf{z}^T\mathbf{Q}_{kl}\mathbf{z}-d \right)^2,~\mathbf{z}^T\mathbf{Q}_{kl}\mathbf{z}-d-\frac{\lambda_{l,k}^{[j]}}{\mu^{[j]}}\leq0,\\-\frac{1}{2\mu^{[j]}}(\lambda_{l,k}^{[j]})^2,~\rm{otherwise}.\quad\quad\quad\quad
            \end{array}\right.
            \\ \label{4-15}
             &v_k^{[j]}\left(\|\mathbf{E}_k(\mathbf{z}-\mathbf{z}_0)\|_2^2-\epsilon^2+t_k\right)+\frac{\mu^{[j]}}{2}\left(\|\mathbf{E}_k(\mathbf{z}-\mathbf{z}_0)\|_2^2+t_k-\epsilon^2\right)^2\notag\\
     &~~=\left\{\begin{array}{l}
        v_k^{[j]}\left(\|\mathbf{E}_k(\mathbf{z}-\mathbf{z}_0)\|_2^2-\epsilon^2\right)+\frac{\mu^{[j]}}{2}\left(\|\mathbf{E}_k(\mathbf{z}-\mathbf{z}_0)\|_2^2-\epsilon^2\right)^2,~-\|\mathbf{E}_k(\mathbf{z}-\mathbf{z}_0)\|_2^2+\epsilon^2-\frac{v_k^{[j]}}{\mu^{[j]}}\leq0,  \\-\frac{1}{2\mu^{[j]}}(v_k^{[j]})^2,~\rm{otherwise}.
     \end{array}\right.
    \end{align}

\setcounter{equation}{\value{MYtempeqncnt}}
\hrulefill
\vspace*{4pt}
\end{figure*}

Substituting (\ref{eq38}) (\ref{eq39}) into (\ref{eq34}) yields an equivalent form $L(\mathbf{z},\pmb{\lambda}^{[j]},\mathbf{v}^{[j]},\mu^{[j]})$ without $\mathbf{r},\mathbf{t}$. The substitution terms of (\ref{eq34}) can be expressed as (\ref{4-14}) and (\ref{4-15}), based on which we can simplify $L(\mathbf{z},\pmb{\lambda}^{[j]},\mathbf{v}^{[j]},\mu^{[j]})$ without $\mathbf{r},\mathbf{t}$ as
\setcounter{equation}{41}
\begin{equation}
\begin{split}
    \label{4-16}
    &L(\mathbf{z},\pmb{\lambda}^{[j]},\mathbf{v}^{[j]},\mu^{[j]})\\&~~~=\mathbf{z}^T\mathbf{z}+\sum_{l=1}^{M}\sum_{k=1,k\neq l}^{M} \phi\left(-\mathbf{z}^T\mathbf{Q}_{kl}\mathbf{z}+d,\lambda_{l,k}^{[j]},\mu^{[j]}\right)\\&~~~~~~~~~~~~+\sum_{k=1}^{M}\phi\left(\|\mathbf{E}_k(\mathbf{z}-\mathbf{z}_0)\|_2^2-\epsilon^2,v_k^{[j]},\mu^{[j]}\right),
    \end{split}
\end{equation}
where the function $\phi(a,b,c)$ is defined as
\begin{equation}
    \label{4-17}
    \phi(a,b,c)\triangleq\left\{\begin{array}{l}
       ba+\frac{c}{2}a^2,~-a-\frac{b}{c}\leq0,\\-\frac{1}{2c}b^2,~\rm{otherwise}.
    \end{array}\right.
\end{equation}

Then, the problem in (\ref{4-11}) becomes finding $\mathbf{z}$ such that $\mathbf{z}^{[j+1]}=\arg \min_\mathbf{z}L(\mathbf{z},\pmb{\lambda}^{[j]},\mathbf{v}^{[j]},\mu^{[j]})$. Since the closed-form solution is difficult to obtain, we use the Broyden–Fletcher–Goldfarb–Shanno (BFGS) algorithm in \cite{Liu1989OnTL} to solve it \footnote{{ The BFGS method is a quasi-Newton method used for solving unconstrained optimization problems. It utilizes the BFGS formula to update the Hessian matrix. Instead of recalculating the Hessian matrix at each iteration, the BFGS method approximates the inverse of the Hessian matrix, thereby avoiding complex computations \cite{Liu1989OnTL}. This algorithm can be implemented in MATLAB using the `fminunc` function from the Optimization Toolbox.}}.

\subsubsection{Update dual variables $\pmb{\lambda}$ and $\mathbf{v}$}
Once the approximate solution $\mathbf{z}^{[j+1]}$ is obtained, according to the Karush–Kuhn–Tucker (KKT) conditions of the augmented Lagrangian \cite{8186925}, the dual variables and the penalty parameters can be updated using the following formulas
\begin{equation}\label{eq44}
    \lambda_{l,k}^{[j+1]}=\left[\lambda_{l,k}^{[j]}-\mu^{[j]}\left((\mathbf{z}^{[j+1]})^T\mathbf{Q}_{kl}\mathbf{z}^{[j+1]}-d\right)\right]^+,
\end{equation}
\begin{equation}\label{eq45}
    v_{k}^{[j+1]}=\left[v_{k}^{[j]}+\mu^{[j]}\left(\|\mathbf{E}_k(\mathbf{z}-\mathbf{z}_0)\|_2^2-\epsilon^2\right)\right]^+,
\end{equation}
and
\begin{equation}\label{eq46}
    \mu^{[j+1]}=\rho\mu^{[j]},
\end{equation}
where $\rho>0$ is the parameter controlling the growth of the penalty parameter.
The detailed steps of the algorithm are summarized in Algorithm 1. It is worth noting that  the result $\mathbf{z}$ from Algorithm 1 will be scaled to meet the total
power constraint at the transmitter.
\renewcommand{\algorithmicrequire}{\textbf{Initialization:}}
\renewcommand{\algorithmicensure}{\textbf{Until:}}
\begin{algorithm}[t]
\caption{ALDA Algorithm for Solving (\textbf{P8}).}
\label{alg:B}
\begin{algorithmic}
    \REQUIRE Initialize $\mathbf{z}^0=\mathbf{1}_{MN} ,  \lambda_{l,k}^0=0.5,\forall l, k, l \neq k, v_k^0=0.5,\forall k,\mu^0=10, \rho=2, $ and $j=0.$
    \REPEAT
    \STATE 1.Update the primal vector $\mathbf{z}^{[j+1]}$ by BFGS in \cite{8267071}
    \begin{equation}
        \mathbf{z}^{[j+1]}=\arg \min_\mathbf{z}L(\mathbf{z}^[j],\pmb{\lambda}^{[j]},\mathbf{v}^{[j]},\mu^{[j]}).\notag
    \end{equation}
    
    \STATE 2. Update the dual vector $\lambda_{l,k}^{[j+1]}$ and $v_k^{j+1}$ through (\ref{eq44}) and (\ref{eq45}), respectively.
    \STATE 3. Update the penalty parameter through (\ref{eq46}).
    \STATE 4. $j=j+1$.

\UNTIL convergence criterion is met.
\end{algorithmic}
\end{algorithm}

\subsection{Complexity Analysis for ALDA}
The main complexity comes from two aspects: 1) the computational complexity of ALDA algorithm at each iteration, and 2) the number of iterations required for the convergence denoted as $K_{ALDA}$. The complexity of each iteration mainly comes from the BFGS solving process and the update of the dual variables. The resulting complexity of the BFGS algorithm for a variable dimension of $MN$ is $O(M^2N^2K_{BFGS})$ with $K_{BFGS}$ being the number of iterations. The resulting complexity of the update of the dual variable is $O(M^4N^2)+O(M^3N^2)$. The total complexity is $O(M^2N^2K_{BFGS}K_{ALDA})+O(M^4N^2K_{ALDA})+O(M^3N^2K_{ALDA})$.

\section{Low-Complexity Solution with BDPS}
The computational complexity associated with ADLA is notably high, particularly in the context of adaptive signal shaping that depends on CSIT.  To reduce the computational complexity, we consider decomposing the original problem (\textbf{P5}) by BDPS. In detail, we split $N$ signal dimensions into $G$ groups, i.e., $\mathcal{N}_1,\mathcal{N}_2,\cdots, \mathcal{N}_G$, to carry $m_1,m_2,\cdots,m_G$ bits subjecting to separate power constraints $P_1, P_2,\cdots, P_G$ and  separate sensing constraints  $\epsilon_1, \epsilon_2,\cdots, \epsilon_G$. 
The separate constraints are limited by the overall constraints, which can be expressed as
\begin{equation}
 P=\sum_{g=1}^G P_g,
 \end{equation}
 and 
 \begin{equation}
 \epsilon^2=\sum_{g=1}^G \epsilon_g^2.
 \end{equation}

 Let $\mathcal{S}_g$ represent the signal set of the $g$th signal dimension group and $M_g$ represent its set sizes  with $M_g=|\mathcal{S}_g|=2^{m_g}$, $\forall g=1,2,\cdots,G$, we have the following relationships
\begin{equation}
M=\prod_{g=1}^{G}M_g,
\end{equation} 
and
\begin{equation}
m=\sum_{g=1}^Gm_g.
\end{equation}

Let $\mathbf{v}_g$ represent the splitting signal vector whose elements are extracted from $\mathbf{s}_0$ corresponding to $g$th signal dimension group. Then, we can formulate small-size problems of finding $\mathcal{S}_g$ given $M_g$, $N_g=|\mathcal{N}_g|$, $\epsilon_g$, $\mathbf{v}_g$, and $\pmb{\Lambda}_g$, where $\pmb{\Lambda}^g\in\mathbb{R}^{2N_r\times N_g}$ is a matrix comprised of the $N_g$ columns corresponding the signal dimension group $\mathcal{N}_g$.  These small-size sub-problems can be transformed similarly to QCQP problems like (\textbf{P5}) and solved by the proposed ALDA algorithm.  The problem size is greatly reduced, leading to computational complexity reduction. Besides,  these subproblems can be solved in parallel, leading to a further reduction of computation time.


Based on the aforementioned BDPS,
the objective of (\textbf{P5}) can be written as
\begin{equation}
\min_{k,l\in [1,M],k\neq l} ||\pmb{\Lambda}(\mathbf{s}_k-\mathbf{s}_l)||_2=\sqrt{\sum_{g=1}^Gd_g^2}
\end{equation} 
where $d_g$ denotes the minimum distance of signals in the $g$th signal dimension group and can be obtained by solving the $g$th subproblem. Then, the optimization can be expressed as
\begin{align}
(\textbf{P10}):~\mathrm{Given}: &~M,N,P,\epsilon,\mathbf{s}_0,\pmb{\Lambda}\notag\\
\mathrm{Find}:&~M_1,M_2,\cdots,M_G,\notag \\&~N_1,N_2,\cdots,N_G,\notag\\&~P_1,P_2,\cdots,P_G,\notag\\&~\epsilon_1,\epsilon_2,\cdots,\epsilon_G,\notag\\
\mathrm{Maximize}:&~ \sqrt{\sum_{g=1}^G d_g^2}\\
\mathrm{Subject~to}:&~\prod_{g=1}^G M_g=M,~\sum_{g=1}^{G}N_g=N\notag,\\
&~\sum_{g=1}^G P_g\leq P\notag,~\sum_{g=1}^G \epsilon_g^2=\epsilon^2\notag.
\end{align}
It is worth mentioning that (\textbf{P10}) is a mixed integer programming and its objective function involves solving a sequence of QCQP problems. {To deal with it, we resort to heuristic algorithms such as genetic algorithm (GA) \cite{10.5555/534133}, and particle swarm optimization (PSO) algorithm \cite{488968}}.
{
Let $N_{iter}$ represent the number of iterations and $N_{pop}$ represent the population size in GA or PSO algorithms. The computational complexity is approximately $O(N_{iter}N_{pop}\sum_{g=1}^G M_g^4 N_g^2)$. Compared to the computational complexity of directly solving (\textbf{P5}), this is significantly reduced. Our simulation results (shown in Section VI-A, Fig. 3) indicate that with $N_{pop} \leq 10$, the algorithm can converge within 10 iterations. Assuming an average of $N_{iter} = 5$, the computational complexity for different parameters and solutions is illustrated in Table \ref{t}. As expected, increasing the number of transmit antennas and the size of the signal set leads to increased complexity. Additionally, the proposed ALDA reduces complexity compared to SDR, and BDPS further reduces it.}
{
\begin{remark}
    Although BDPS can decompose a large-scale QCQP into a sequence of sub-QCQP problems, which indeed reduces the complexity of the QCQP problem, it also introduces the issue of mixed integer programming, i.e., how to divide \( N \) into \( G \) groups. This brings additional computational overhead.
    As a result, there is a trade-off in the BDPS algorithm between the additional computational overhead and the reduction in complexity. When the values of \( N \) and \( M \) are significantly large, the complexity reduction brought by BDPS through dimensionality reduction of the QCQP problem far outweighs the additional computational overhead from the mixed integer programming problem. Conversely, when the problem size is not very large, the additional overhead brought by BDPS will increase the problem's complexity. Therefore, BDPS is a low-complexity algorithm suitable for large-scale problems.
\end{remark}
}

\begin{table}[t]
\centering
\caption{{Complexity Comparison.\\$(32,8,4)$ means $N/2=32, N_r=8, M=4$.}}
\label{t}
\renewcommand{\arraystretch}{1.5}
\begin{tabular}{c|c|c|c}
\hline
\textbf{Method} &$(16,8,4)$  &$(32,8,4) $& $(32,8,16)$ \\ \hline
SDR      &       $ 2.37\times 10^7$  &  $ 2.68\times 10^8$& $3.04\times 10^9 $                           \\ \hline
ALDA       &       $ 6.55\times 10^5$  &  $ 2.62\times 10^6$& $4.19\times 10^7$                               \\ \hline
BDPS      &       $1.02\times 10^5$  &  $ 4.10\times 10^5$& $1.06\times 10^6 $                                \\ \hline
\end{tabular}
\end{table}

\section{Simulation Results and Discussions}
In this section, we provide thorough comparisons with existing approaches in ISAC signal design. As far as we are aware, our work is the first to address the generalized problem of ISAC signal design. i.e., (\textbf{P2}). For comparison,
we choose existing works on precoding/beamforming for ISAC because all modifications on the precoder/beamformer will affect the signal set as
\begin{equation}
\mathbf{x}_i=\mathbf{F}\mathbf{s}_i,
\end{equation}
where $\mathbf{F}$ can be regarded as a beamforming matrix (in the spatial domain) or a precoding matrix (in the time/frequency domain). 

{
Specifically, we compare the existing signal designs that optimize the beamformer/precoder in MIMO scenarios, as the signal design considering the ICI is the most complicated. The following schemes are chosen for comparison.

\begin{itemize}
    
    \item The symbol-level ISAC beamforming with interference reduction (SL-IR) in \cite{8386661,2021JointTW}, focuses on reducing interference at the symbol level. Symbol-level precoding is designed for communication systems with finite constellation inputs and can effectively eliminate ICI impacts on the transmitted symbols.
    \item The block-level ISAC beamforming for maximizing minimum Euclidean distance (BL-MMED) with finite alphabet input in \cite{9850347}. This method is a low-complexity precoding approach tailored for communication systems with  finite alphabet. Compared to SL-IR, it not only considers the practical requirements of communication systems but also takes complexity into account.
    \item The block-level ISAC beamforming for Gaussian inputs (BL-Gaussian) in \cite{8683591,10.1145/3556562.3558569}. This method is widely adopted in most current ISAC research. It does not consider the transmission requirements of practical communication systems and is based on Shannon theory to maximize the spectral efficiency. 
\end{itemize}

In summary, the three methods mentioned above are the most common approaches in ISAC signal design for MIMO systems. They encompass different communication input assumptions and various levels of precoding.}
\begin{figure}[!t]
       \centering
       \includegraphics[width=0.95\linewidth]{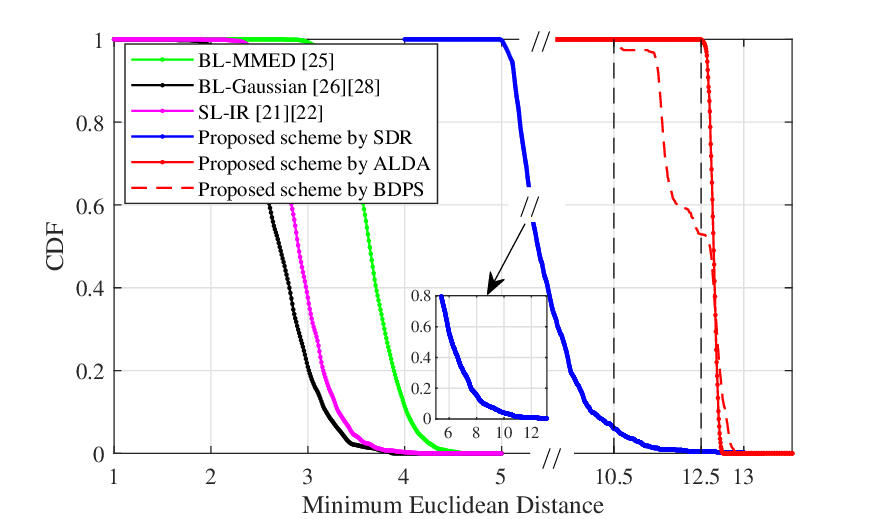}
       \caption{CDF of the minimum Euclidean distances among the noise-free received vectors for different schemes, where $d=14$ and $\epsilon=0.3$.}
       \label{fig2}    
\end{figure}
In our simulation, we configure the carrier frequency at $24$ GHz. The transmitter is equipped with $N/2 = 32$ antennas, while the receiver has $N_r = 8$ antennas. The transmission power is set at $P = 1$. The signal set size is chosen as $M=4$. We model the downlink channel $\mathbf{H}$ as a flat Rayleigh fading channel, where each element of the channel matrix $\mathbf{H}$ follows a standard Complex Gaussian distribution. The CSIT is assumed to be perfectly known. 
 {
Furthermore, we utilize spatial waveforms as reference sensing waveforms. For search tasks, the reference waveform is designed to scan a wide main lobe spanning \([-20^\circ, 20^\circ]\) with sidelobe levels (SLL) kept below 15 dB. For tracking tasks, the reference waveform is designed to produce a narrow beam focusing on a target located at $15^\circ$.
The spatial reference waveforms can be obtained by solving the following problem, as detailed in literature \cite{7347464}.
\begin{align}
(\textbf{Wide beam}):~~&\min_{\mathbf{x}_0}~\max_{\theta_i}|P_d(\theta_i)-\mathbf{x}_0^H\mathbf{a}(\theta_i)|,~\theta_i\in \mathbf{\Theta}_m\notag\\ 
&~~~\text{s.t.}~\max_{\theta_i}|\mathbf{x}_0^H\mathbf{a}(\theta_j)|\leq\gamma,\theta_j\in \mathbf{\Theta}_s,
\end{align}
\begin{align}
(\textbf{Narrow beam}):~~&\min_{\mathbf{x}_0}~\max_{\theta_j}|\mathbf{x}_0^H\mathbf{a}(\theta_j)|,~\theta_j\in \mathbf{\Theta}_s\notag\\ 
&~~~\text{s.t.}~|\mathbf{x}_0^H\mathbf{a}(\theta_t)|=1,
\end{align}
where $P_d(\theta_i)$ is the desired transmit beampattern, $\mathbf{\Theta}_m$ is the wide main-lobe spatial sector, $\mathbf{\Theta}_m$ is the wide main-lobe spatial sector, $\mathbf{\Theta}_s$ is the side-lobe sector, $\gamma$ is the desired SLL, $\theta_t$ is the direction of narrow beam.}

These simulation parameters will remain constant unless stated otherwise.

\begin{figure}[!t]
       \centering
       \includegraphics[width=0.95\linewidth]{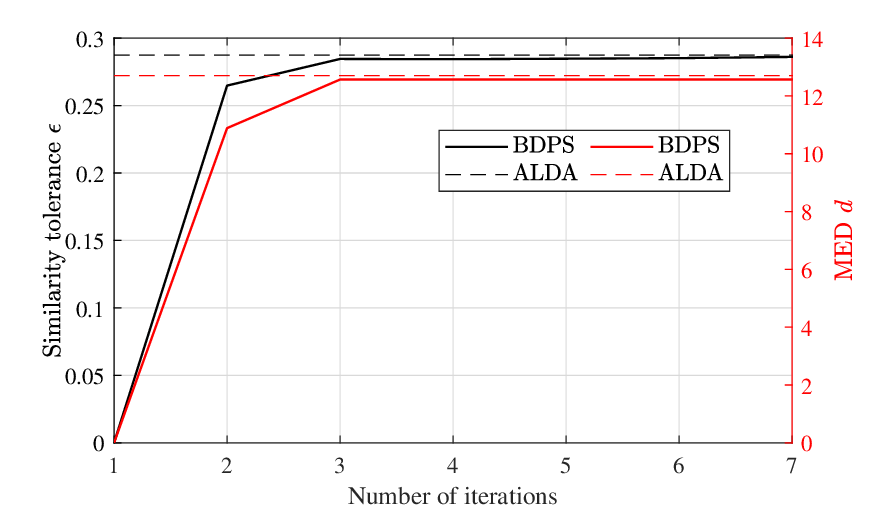}
       \caption{Coverage performance of the proposed BDPS.}
       \label{fig3}    
\end{figure}

\subsection{Communications Performance Evaluation}
We conducted an evaluation of the minimum Euclidean distance (MED) performance across various signal design strategies. Over 1000 channel realizations, we plotted the cumulative distribution function (CDF) of the minimum Euclidean distances among all noise-free received vectors for different schemes and algorithms, as shown in Fig. \ref{fig2}, where the minimum distance $d=14$ in (\textbf{P8}) and $\epsilon=0.3$. The results demonstrate that our proposed signal design scheme outperforms others, consistently achieving a minimum Euclidean distance greater than $12.5$. The steep curve corresponding to the ALDA algorithm signifies its stability. In contrast, the minimum Euclidean distances for other schemes like BL-Gaussian, BL-MMED, and SL-IR barely reach $5$, with the probability Pr$\{d>5\}$ being almost zero.
This performance gap is attributed to their constrained modulation schemes, which hinder reaching a globally optimal solution. Our proposed approach, treating ISAC signal as a high-dimensional modulation signal design problem, effectively overcomes this limitation. When compared to traditional SDR-solving methods, it's observable that, despite SDR with randomization providing superior solutions than other schemes, they lack the stability of those achieved by ALDA. This highlights the performance limitations of the SDR algorithm in large-scale QCQP problems \cite{8267071}.
Moreover, as a complexity-reducing algorithm, BDPS closely approaches ALDA’s performance. Fig. \ref{fig3} illustrates the convergence curve of the BDPS algorithm, showing rapid convergence within a few iterations.

\begin{figure}[!t]
       \centering
       \includegraphics[width=0.95\linewidth]{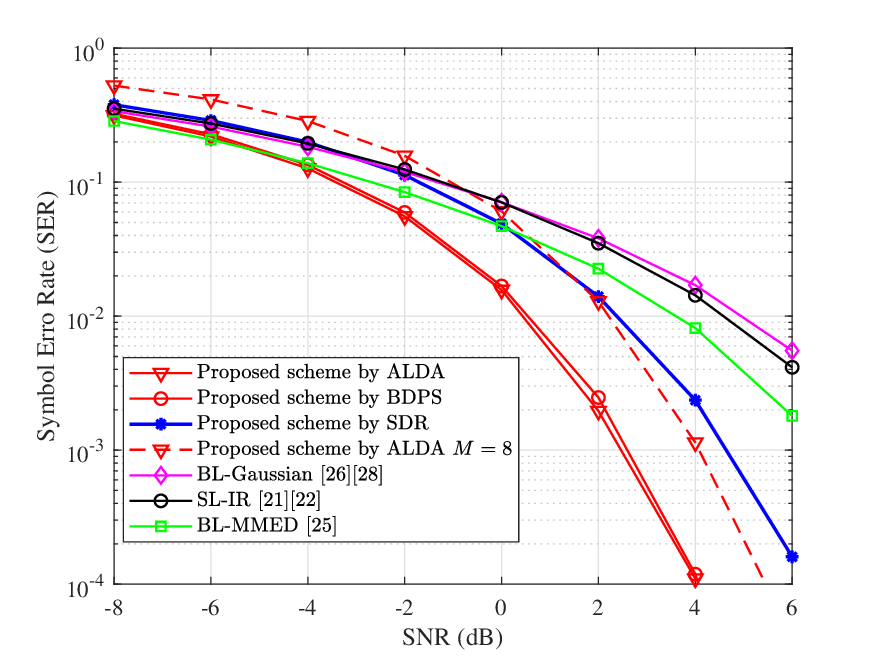}
       \caption{SER of different signal design schemes.}
       \label{fig4}    
\end{figure}
\begin{figure}[!t]
       \centering
       \includegraphics[width=0.95\linewidth]{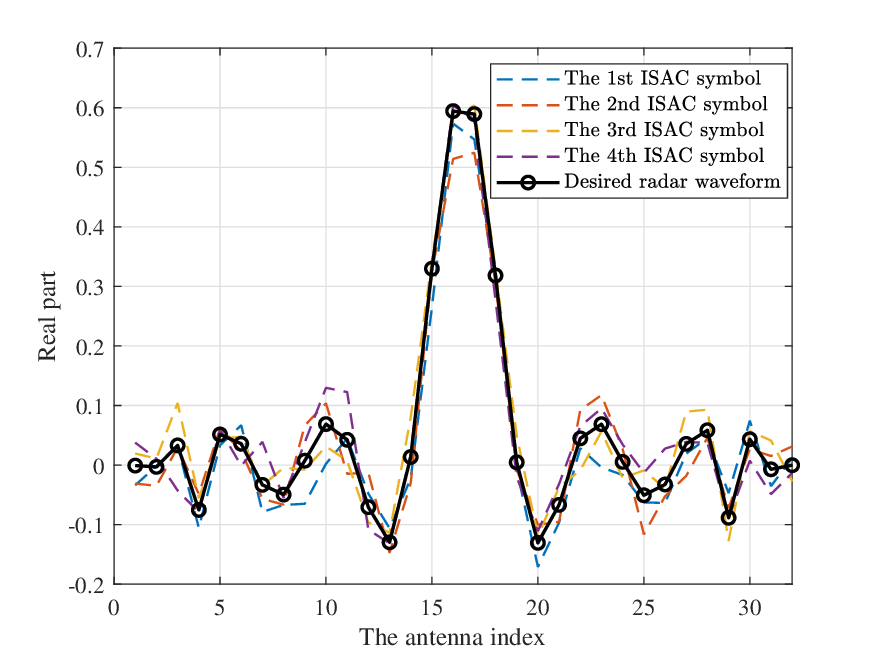}
       \caption{ISAC symbols designed using the proposed scheme.}
       \label{fig5}    
\end{figure}
The Euclidean distance is a critical metric in our study as it signifies the distinguishability between different symbols and is directly correlated with the SER. A larger Euclidean distance typically results in a lower SER. In Fig. \ref{fig4}, we present the SER performance of various design schemes. The findings echo our previous conclusions. At $4$ dB, the SER for our proposed design scheme reaches as low as $10^{-4}$, which is notably lower by two orders of magnitude compared to benchmark schemes. The proposed BDPS approach demonstrates SER performance comparable to the ALDA approach. Furthermore, we extended our analysis to simulate the SER curve for a signal set size of $M=8$. It was observed that increasing the signal set size adversely affects the SER. This deterioration is attributed to the reduction in the minimum Euclidean distance under the constant similarity constraints of the sensing waveform, which occurs as the number of ISAC symbols increases. Despite the higher transmission rate, our proposed scheme maintains superior performance over comparative algorithms at an SNR greater than $2$ dB.

\subsection{Sensing Performance Evaluation}

\begin{figure}[!t]
       \centering
       \includegraphics[width=0.95\linewidth]{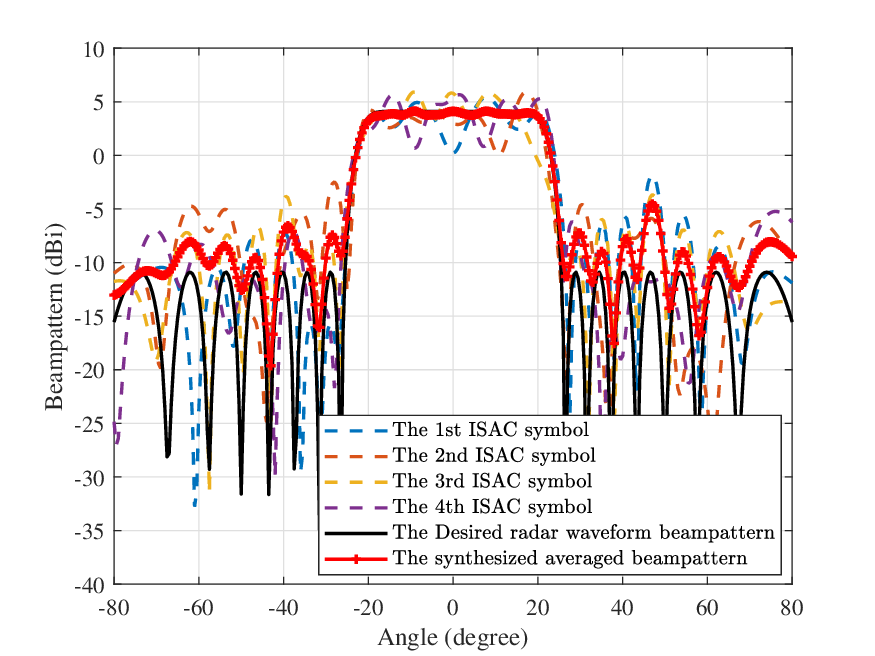}
       \caption{Beampattern of ISAC symbols designed using the proposed scheme.}
       \label{fig6}    
\end{figure}

\begin{figure}[!t]
       \centering
       \includegraphics[width=0.95\linewidth]{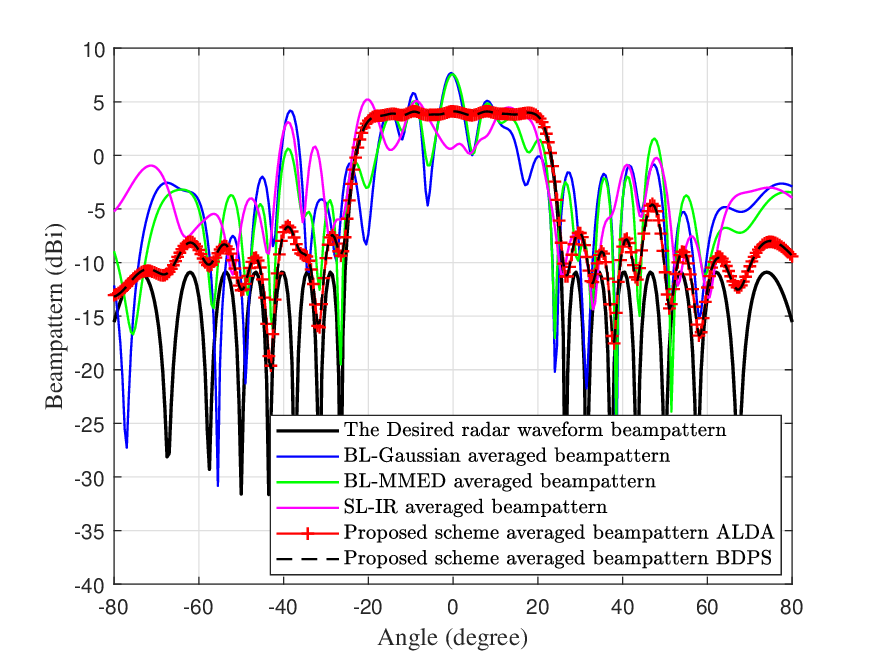}
       \caption{Average beampatterns of different schemes.}
       \label{fig7}    
\end{figure}

Secondly, we assess the sensing performance through the analysis of waveforms and beampatterns of the ISAC signals, as illustrated in Figs. \ref{fig5} and \ref{fig6}. The proposed scheme's ISAC symbols, as seen in Fig. \ref{fig5}, adhere closely to the reference signal under similarity constraints. This indicates that transmitting any ISAC symbol from the designed set ensures robust communication while satisfying sensing waveform requirements.

In the spatial domain, as observed in Fig. \ref{fig6}, the ISAC symbols align well with the main lobe range. By transmitting each ISAC symbol an equal number of times within a communication frame, we plotted the average beampattern over this period, shown as the red line in the figure. This average beampattern stabilizes and closely approximates the reference beampattern after signal accumulation. Fig. \ref{fig7} contrasts this with the average beampatterns of other schemes, highlighting the superiority of our proposed approach.

\begin{figure}[!t]
       \centering
       \includegraphics[width=0.95\linewidth]{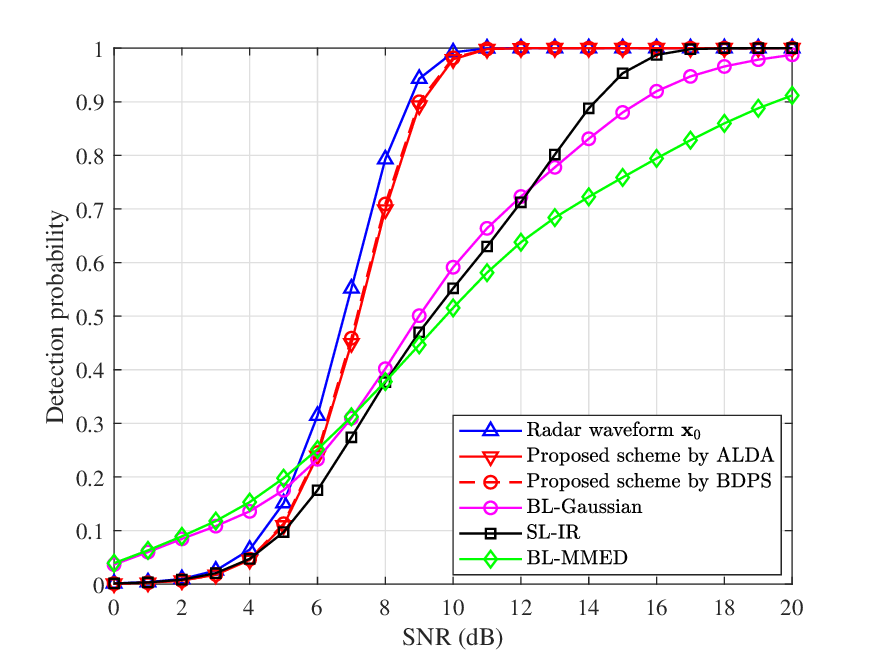}
       \caption{Detection probability comparisons among different schemes.}
       \label{fig8}    
\end{figure}

Moreover, wide beams, vital for scanning tasks, are evaluated using detection probability as a performance metric. Assuming a far-field target, we conducted $1000$ Monte Carlo simulations with the target's angle appearing randomly within the main lobe range of $[-20^\circ, 20^\circ]$. The results, depicted in Fig. \ref{fig8}, reveal that the proposed scheme's curve closely matches that of the reference sensing waveform, owing to its high waveform similarity {(i.e. low similarity tolerance $\epsilon$)}. Notably, the ALDA and BDPS schemes under our design achieve a significant gain of at least $6$ dB in detection probability compared to other schemes, affirming their effectiveness at a detection probability of $1$.

\begin{figure}[!t]
       \centering
       \includegraphics[width=0.95\linewidth]{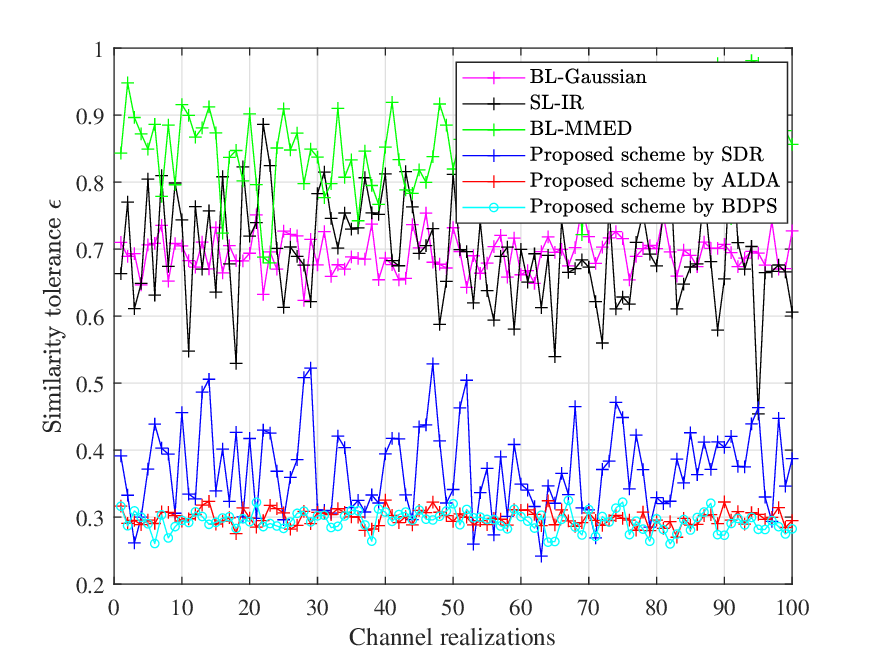}
       \caption{Waveform similarity of different schemes under different channel realizations.}
       \label{fig9}    
\end{figure}
To further examine the influence of different channel conditions on the sensing performance of ISAC signals, we graphed the waveform similarity across various channel realizations, as depicted in Fig. \ref{fig9}. The data shows that the waveform similarity for the benchmark algorithms consistently remained above $0.6$, yet these values exhibited significant fluctuations in response to channel changes. Such variability implies that radar performance is highly sensitive to channel variations, leading to potential instability which is undesirable in radar applications. A similar issue is observed with the SDR-solving approach.
In contrast, the waveform similarity for the proposed ALDA scheme demonstrates remarkable stability, maintaining around $0.3$ with minimal fluctuation across $100$ channel realizations. This consistency underscores the exceptional stability of the sensing performance of our proposed scheme. In the face of dynamically changing communication channels, our ISAC signal design adeptly fulfills sensing requirements, showcasing outstanding robustness and adaptability to varying channel conditions.

\begin{figure}[!t]
       \centering
       \includegraphics[width=0.95\linewidth]{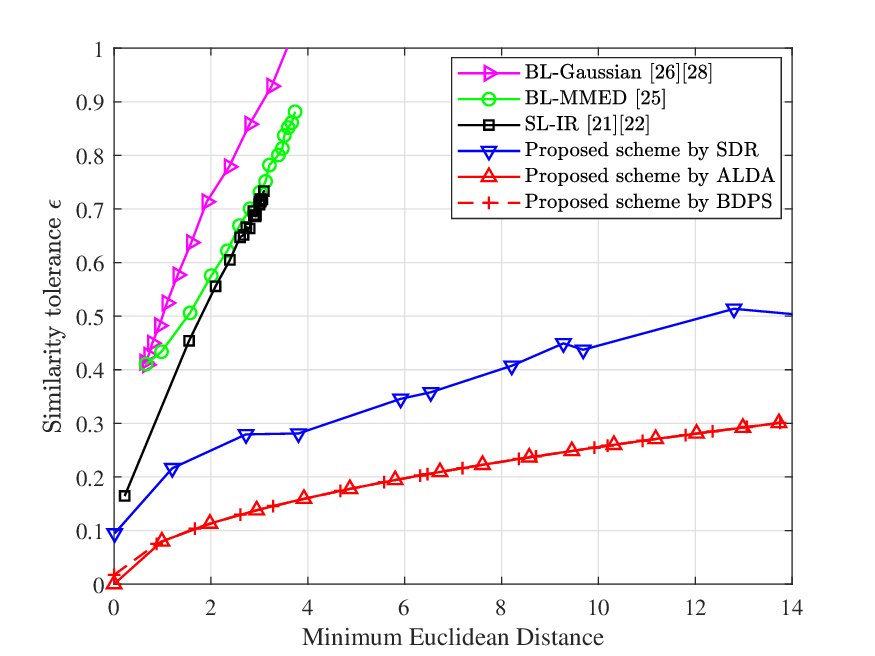}
       \caption{Trade-off of the similarity and the minimum Euclidean distance.}
       \label{fig10}    
\end{figure}
\begin{figure}[!t]
       \centering
       \includegraphics[width=0.95\linewidth]{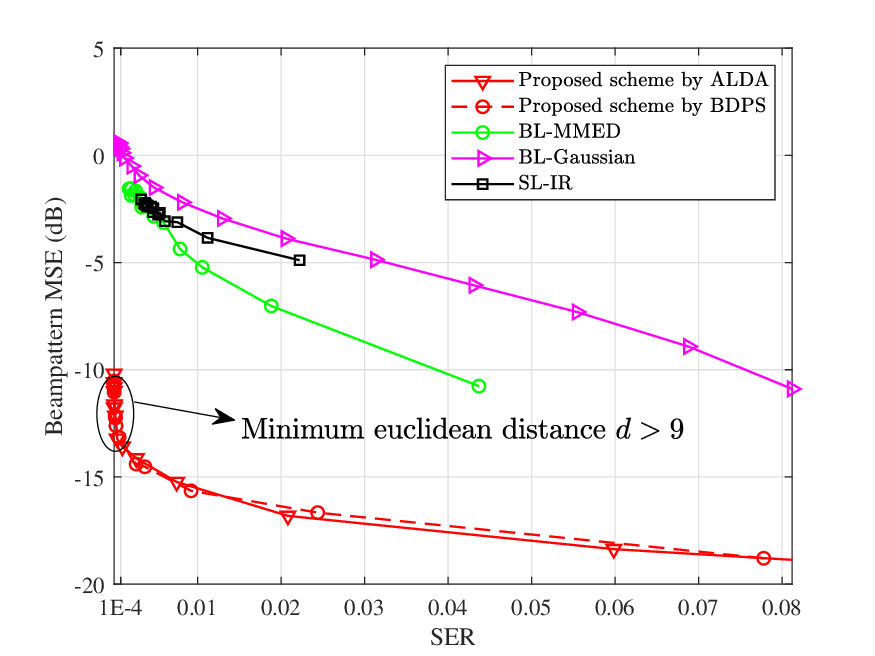}
       \caption{Trade-off of the beampattern MSE and the SER, SNR = $6$ dB.}
       \label{fig11}    
\end{figure}
\subsection{Performance Trade-off Between Sensing and
Communications}
Our proposed scheme aims to identify ISAC signals that fulfill a set minimum Euclidean distance within a space constrained by sensing waveform similarity. This approach allows for a balanced performance trade-off between communication and sensing by adjusting the minimum Euclidean distance $d$ or the similarity constraint $\epsilon$. In our simulation, we kept the similarity constraint constant at $\epsilon=0.3$ while varying $d$ from 0 to 14. For comparison, we adjusted the trade-off parameter of the benchmark algorithms from $0.01$ to $0.99$.

The results of this performance trade-off are illustrated in Fig. \ref{fig10}. We observe that an increase in the minimum Euclidean distance correlates with a higher tolerance in similarity. As hypothesized, the proposed ALDA and BDPS algorithms substantially outperform other schemes. Our proposed scheme consistently exhibits a significantly lower signal similarity tolerance for any given minimum Euclidean distance. The smoothness of our scheme's curves reflects its effective balance between communication and sensing.
To provide further insight, we plotted additional curves in Fig. \ref{fig11}, using beampattern MSE and SER as metrics. The majority of the points on the ALDA and BDPS curves fall within a delineated black ellipse region. In this zone, both a SER below $10^{-4}$ and a beampattern MSE under $-10$ dB are attainable, further evidencing the proposed scheme’s ability to achieve an optimal compromise between reliable communication and accurate sensing. 
\begin{figure}[!t]
       \centering
       \includegraphics[width=0.95\linewidth]{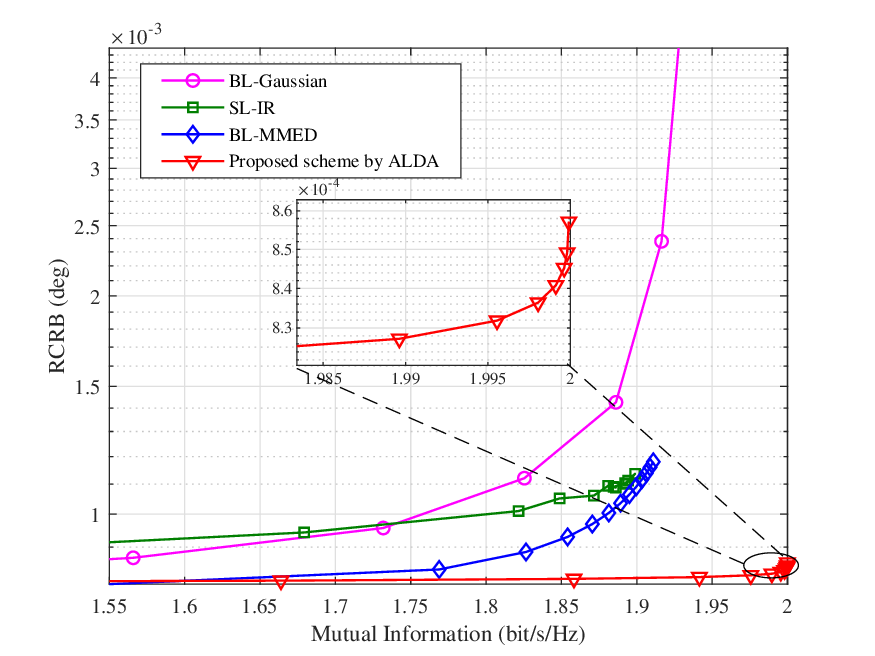}
       \caption{Trade-off of the sensing CRB and communication MI, SNR = $0$ dB.}
       \label{fig11_b}    
\end{figure}
{
\begin{figure}[!t]
       \centering
       \includegraphics[width=0.95\linewidth]{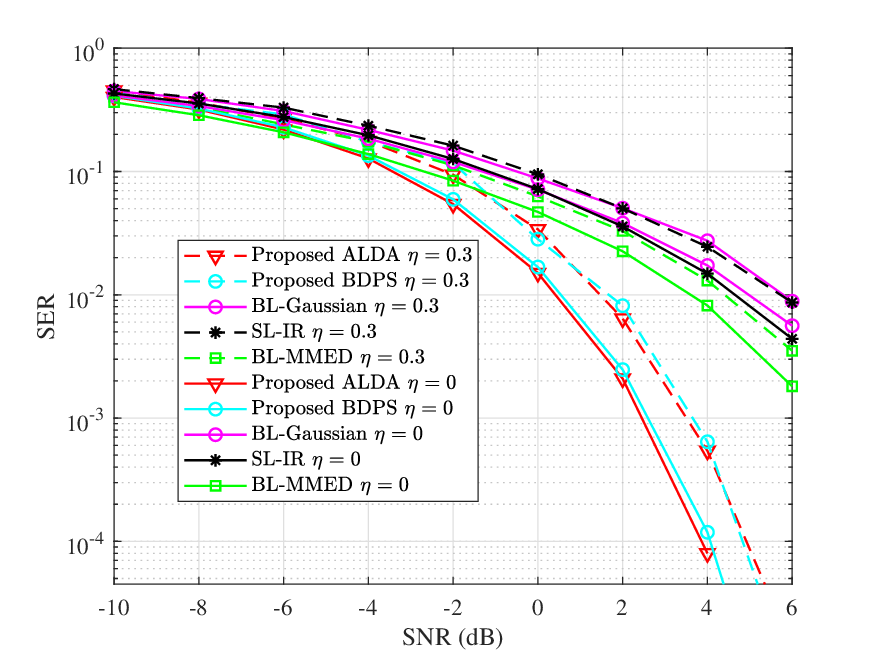}
       \caption{SER of different signal design schemes under imperfect CSIT.}
       \label{fig12}    
\end{figure}

Additionally, the CRB metric for sensing and the communication rate metric are also important standards for evaluating ISAC signals. At this point, the sensing reference signal adopts the narrow beam signal shown in Equation (55), assuming the sensing target is located at \(\theta_t = 0^\circ\). For the CRB calculation method, see [\citenum{9652071}, Eq. (7)] .
 For communication, the communication rate with finite constellation input can be calculated using MI. Given \(\mathcal{X}\) as the input, the detailed calculation method for the MI of the MIMO system can be found in \cite{6107431}. In Fig. \ref{fig11_b}, we show the trade-off between root-CRB (RCRB) and MI for different schemes.
 Since MI is directly related to the MED between received signals, the proposed scheme can achieve maximum MI, i.e. $\log_2M=2$ bit/s/Hz, while maintaining the lowest CRB.

 }


\begin{figure}[!t]
       \centering
       \includegraphics[width=0.95\linewidth]{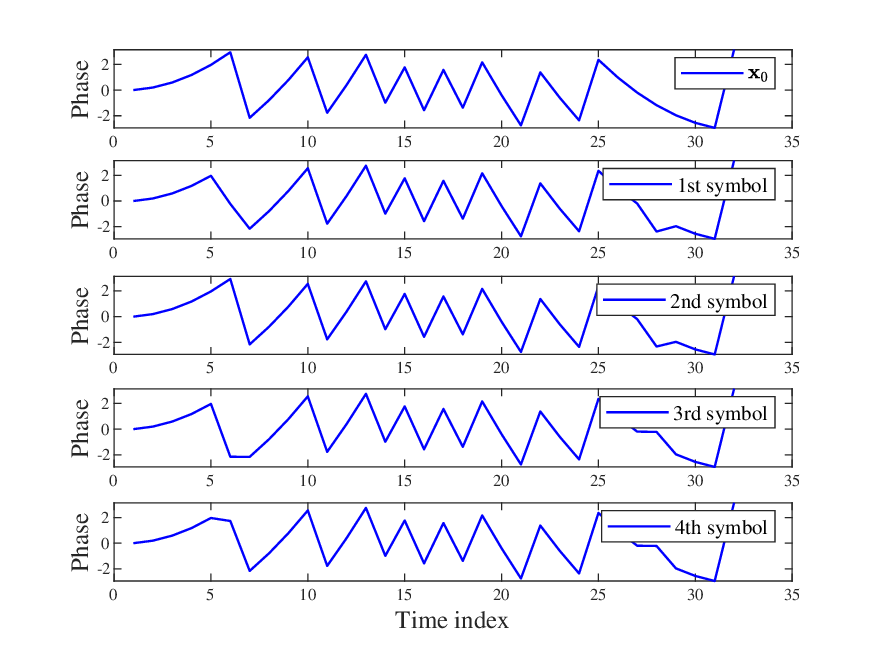}
       \caption{The phase of the time signal obtained by the proposed scheme, where $d=2$ and $\epsilon=0.5$.}
       \label{fig13}    
\end{figure}

\begin{figure}[!t]
       \centering
       \includegraphics[width=0.95\linewidth]{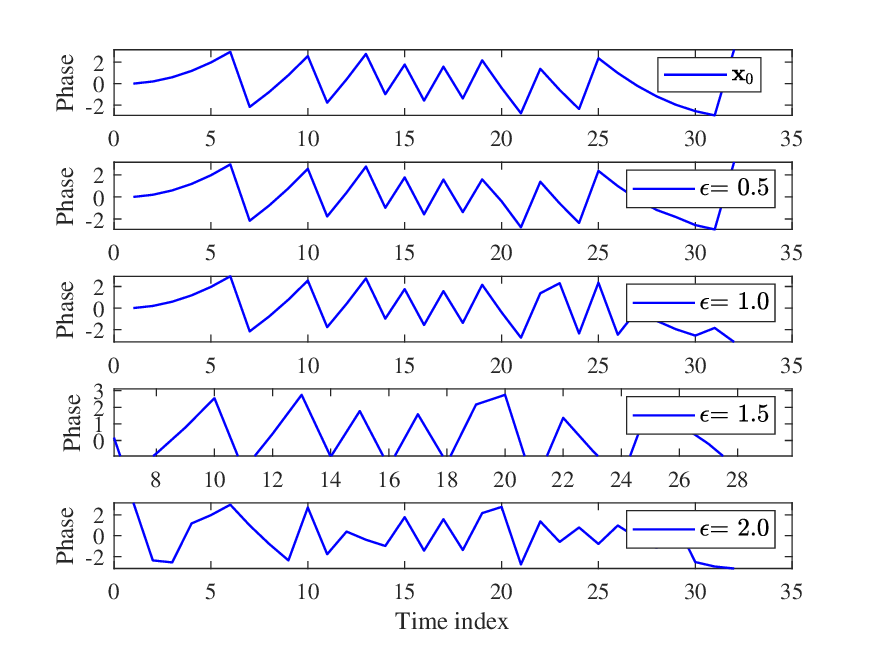}
       \caption{The phase of the time signal obtained by the proposed scheme for different similarity tolerance.}
       \label{fig14}    
\end{figure}
\subsection{Impact of imperfect CSIT on ISAC Signal Performance} 

The signal design schemes discussed thus far are based on the premise of perfect CSIT, a condition that is often unattainable in real-world scenarios. Therefore, it is imperative to assess the signal's robustness in the face of channel estimation errors or imperfect CSIT. We consider the channel error model as defined in \cite{8734877}:
\begin{equation}
\mathbf{H}_{im} = \mathbf{H} + \mathbf{H}_e,
\end{equation}
where $\mathbf{H}_e$ represents the matrix of channel errors, with each element following a complex Gaussian distribution with zero mean and variance $\sigma_e^2$. Assuming the application of Least Squares (LS) channel estimation, the channel estimation error $\sigma_e^2$ is proportional to the noise variance $\sigma_n^2$, specifically, $\sigma_e^2 = \eta\sigma_n^2$. In our simulations, we set $\eta = 0.3$ and compared these results with those obtained under perfect CSIT conditions, as depicted in Fig. \ref{fig12}. The introduction of channel errors leads to a degradation in the SER across all schemes. However, our proposed scheme still surpasses comparative schemes in performance, demonstrating its robustness in communication even under imperfect CSIT.

Furthermore, it is noteworthy that channel estimation errors can be interpreted as channel variations, transitioning from $\mathbf{H}$ to $\mathbf{H}_{im}$. Consequently, as indicated in Fig. \ref{fig9}, our proposed schemes exhibit resilience to such channel variations. This implies that the sensing performance of our proposed scheme remains consistently effective for both $\mathbf{H}$ and $\mathbf{H}_{im}$.

\begin{figure}[!t]
       \centering
       \includegraphics[width=0.95\linewidth]{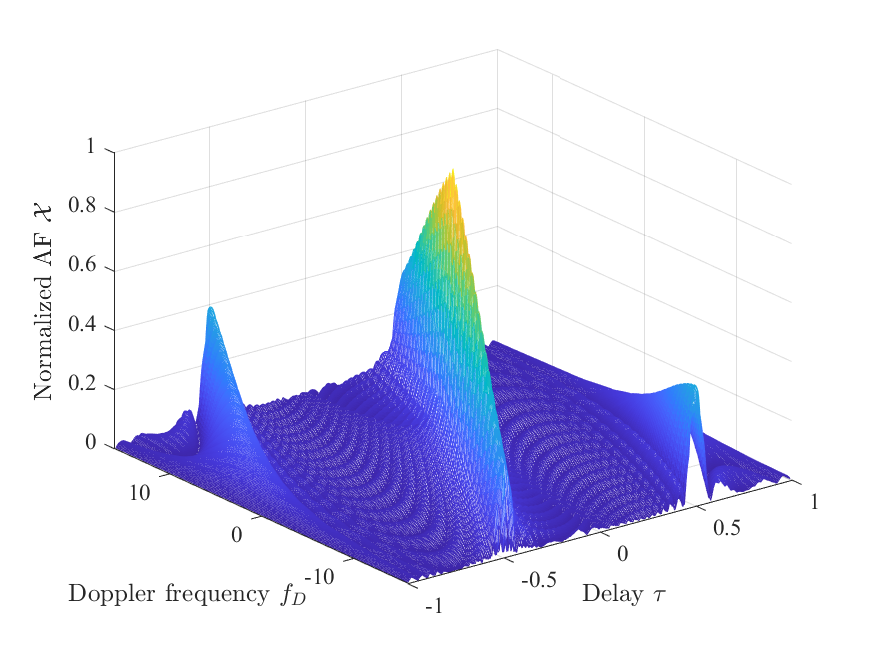}
       \caption{Ambiguity function of the LFM signal.}
       \label{fig15}    
\end{figure}

\subsection{Extension to Time Domain ISAC Signal Set Design}
As outlined in Section II, our research adopts a unified approach to ISAC signal design. In this section, we extend our proposed scheme to the design of time domain signals. We assume that each ISAC symbol is transmitted over $\frac{N}{2} = 32$ time slots. For our sensing reference, we employ the classical LFM signal, renowned for its excellent pulse compression and ambiguity properties, as detailed in \cite{2021JointTW}.
The time sequence of the LFM signal $\mathbf{x}_0$ is given by [\citenum{he_li_stoica_2012}, eq.(1.21)],
\begin{equation}
    x_0(n)=\sqrt{\frac{2P}{N}}e^{j\pi \frac{2n(n-1)}{N}},~n=1,2,\cdots,\frac{N}{2}.
\end{equation}

With the minimum Euclidean distance set to $d=2$ and the similarity constraint at $\epsilon=0.5$, Fig. \ref{fig13} displays the phase of signals generated by our proposed scheme. To assess similarity, we also include the phase of the reference LFM signal. The comparison reveals that each ISAC symbol closely approximates the reference signal, effectively balancing between meeting communication requirements and maintaining similarity.

Subsequently, in Fig. \ref{fig14}, we examine the impact of varying similarity constraints. For simplicity, we illustrate this using just one ISAC signal under each constraint. As anticipated, a smaller $\epsilon$ value implies a stronger similarity constraint, resulting in signals more closely resembling the LFM signal.

\begin{figure}[!t]
       \centering
       \includegraphics[width=0.95\linewidth]{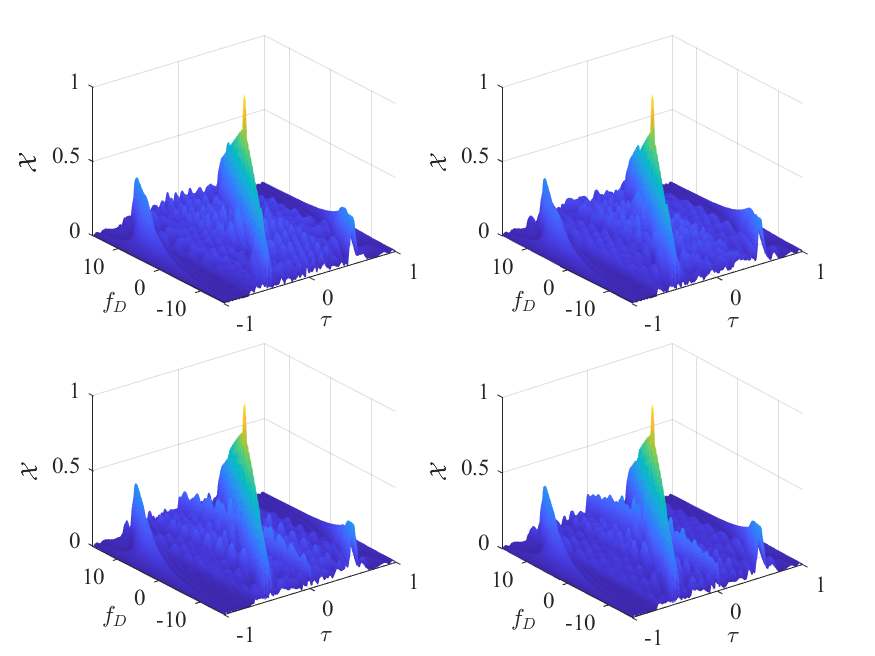}
       \caption{Ambiguity function of the designed ISAC symbols.}
       \label{fig16}    
\end{figure}

Finally, the ambiguity functions of both the LFM and the ISAC signals are presented in Figs. \ref{fig15} and \ref{fig16}. The ambiguity functions of the designed ISAC symbols exhibit a knife-edge shape, which is indicative of their enhanced preservation of the ambiguity properties inherent to the LFM signal.
{
\subsection{Discussion on the applicability of the proposed scheme}
Since the proposed scheme is more generalized, it can be applied to many ISAC-related scenarios. For example, \cite{10217150} proposed an integrated sensing-communication-computation framework, where ISAC is achieved through time-division, using orthogonal time resources for sensing and communication separately. During the sensing period, frequency-modulated continuous-wave (FMCW) waveforms are transmitted, and during the communication period, data modulated with QAM, etc., are transmitted. This approach does not fully utilize time resources. 
Combining our scheme, the ISAC implementation method can be changed from time-division to a fully integrated scheme. Using our proposed signal design method, FMCW can be used as the reference sensing signal, and the ISAC signal set can be optimized accordingly. In this case, the sensing period and the communication period are fully integrated, and the DFRC transmitter emits an integrated signal that performs communication data transmission while simultaneously executing sensing. This achieves complete fusion of communication and sensing, further enhancing system performance.}

\section{Conclusion}
This paper presents a unified approach to the optimization of ISAC signal set design. We developed a unified signal design framework applicable both with and without knowledge of CSIT. This framework encompasses both parallel and cross-talk channel scenarios. We transformed this unified challenge into a large-scale QCQP problem. To address its significant computational complexity, we introduced an efficient ALDA algorithm tailored for large-scale QCQP. Additionally, to further reduce complexity, we proposed a low-complexity problem decomposition method utilizing BDPS. Our simulation results highlight the effectiveness of our proposed signal designs, demonstrating their superiority over existing signal design methods in the ISAC domain.

\bibliographystyle{IEEEtran} 
\bibliography{IEEEabrv,bib}

\begin{IEEEbiography}[{\includegraphics[width=1in,height=1.25in,clip,keepaspectratio]{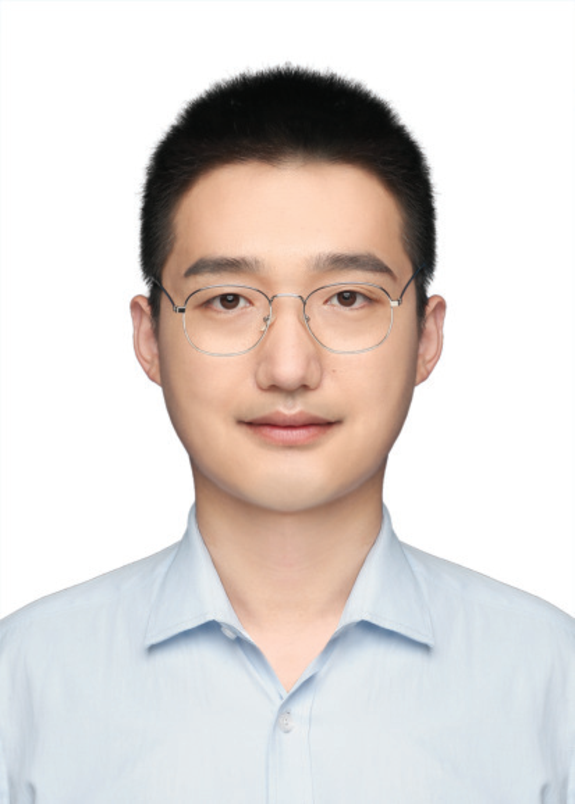}}]{Shuaishuai Guo}(Senior Member, IEEE) received the B.E and Ph.D. degrees in communication and information systems from the School of Information Science and Engineering, Shandong University, Jinan, China, in 2011 and 2017, respectively. He visited University of Tennessee at Chattanooga (UTC), USA, from 2016 to 2017. He worked as a postdoctoral research fellow at King Abdullah University of Science and Technology (KAUST), Saudi Arabia from 2017 to 2019. Now, he is working as a full professor of Shandong University. His research interests include 6G communications and machine learning.
\end{IEEEbiography}

\begin{IEEEbiography}[{\includegraphics[width=1in,height=1.25in,clip,keepaspectratio]{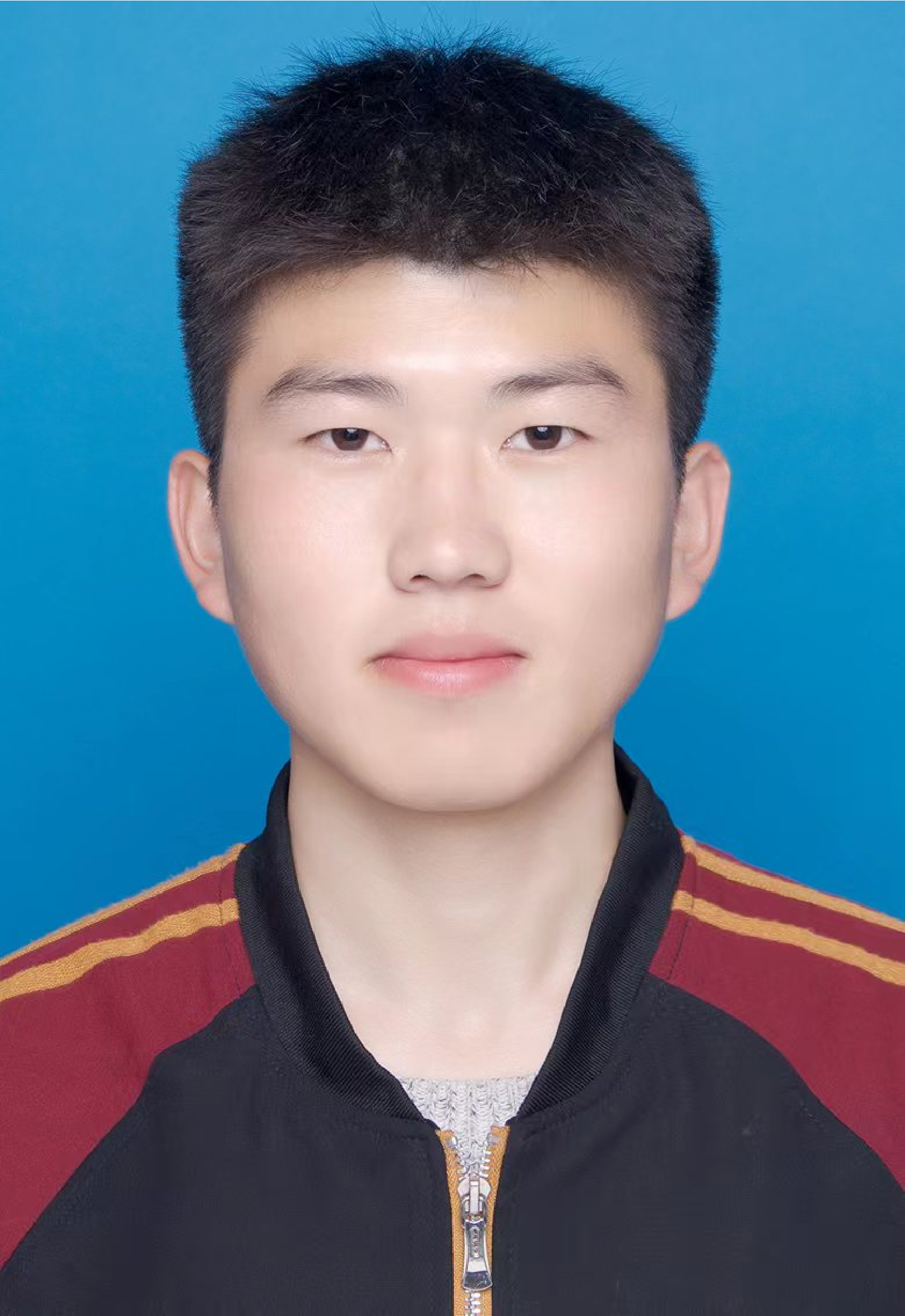}}]{Kaiqian Qu}(Graduate Student Member, IEEE) was born in  Shandong Province, China. He received the B.E. degree from the School of Physics, Zhengzhou University, China, in 2022. Now, he is currently pursuing the M.S.  egree at Shandong University, Jinan, China. His main research interests include the integrated sensing and communication and extremely large-scale MIMO.
\end{IEEEbiography}



\end{document}